%% file: main.tex
\documentclass[sn-mathphys-num]{jnl}

\input{preamble.tex}

\begin{document}


\subfile{sections/abstract}


\subfile{sections/introduction}

\subfile{sections/results}

\subfile{sections/discussion}

\subfile{sections/methods}

\subfile{sections/additional}

\pagebreak
\bibliography{bibliography}

\pagebreak
\subfile{sections/supplemental}

\end{document}

%% file: preamble.tex
\usepackage{amsmath,amssymb,amsfonts}%
\usepackage{amsthm}%

\usepackage{mathtools}

\numberwithin{equation}{subsection}
\renewcommand{\figurename}{Figure}

\usepackage{xr}

\usepackage{subfiles} 

%% file: sections/abstract.tex
\title[Article Title]{Hierarchical Bayesian estimation of motor-evoked potential recruitment curves yields accurate and robust estimates}


\author*[1,2]{\fnm{Vishweshwar} \sur{Tyagi}}\email{vt2353@cumc.columbia.edu}
\author[8,9]{\fnm{Lynda M.} \sur{Murray}}
\author[5]{\fnm{Ahmet S.} \sur{Asan}}
\author[3,6]{\fnm{Christopher} \sur{Mandigo}}
\author[4]{\fnm{Michael S.} \sur{Virk}}
\author[7,8,9]{\fnm{Noam Y.} \sur{Harel}}
\author[1,2,4]{\fnm{Jason B.} \sur{Carmel}}
\author*[1,2,4]{\fnm{James R.} \sur{McIntosh}}\email{jrm2263@cumc.columbia.edu}

\affil*[1]{\orgdiv{Neurology}, \orgname{Columbia University}, \orgaddress{\city{New York}, \state{NY}, \postcode{10032}}}
\affil[2]{\orgdiv{Orthopedic Surgery}, \orgname{Columbia University}, \orgaddress{\city{New York}, \state{NY}, \postcode{10032}}}
\affil[3]{\orgdiv{Neurological Surgery}, \orgname{Columbia University}, \orgaddress{\city{New York}, \state{NY}, \postcode{10032}}}
\affil[4]{\orgdiv{Neurological Surgery}, \orgname{Weill Cornell Medicine}, \orgaddress{\city{New York}, \state{NY}, \postcode{10065}}}
\affil[5]{\orgdiv{Staley Center for Psychiatric Research}, \orgname{Broad Institute of MIT and Harvard}, \orgaddress{\city{Cambridge}, \state{MA}, \postcode{02142}}}
\affil[6]{\orgdiv{New York Presbyterian}, \orgname{The Och Spine Hospital}, \orgaddress{\city{New York}, \state{NY}, \postcode{10034}}}
\affil[7]{\orgdiv{Neurol.}, \orgname{Icahn School of Medicine at Mount Sinai}, \orgaddress{\city{New York}, \state{NY}, \postcode{10029}}}
\affil[8]{\orgdiv{Rehabilitation and Human Performance}, \orgname{Icahn School of Medicine at Mount Sinai}, \orgaddress{\city{New York}, \state{NY}, \postcode{10029}}}
\affil[9]{\orgname{James J. Peters Veterans Affairs Medical Center}, \orgaddress{\city{Bronx}, \postcode{10468}, \state{NY}}}


\abstract{
    Electromagnetic stimulation probes and modulates the neural systems that control movement. Key to understanding their effects is the muscle recruitment curve, which maps evoked potential size against stimulation intensity. Current methods to estimate curve parameters require large samples; however, obtaining these is often impractical due to experimental constraints. Here, we present a hierarchical Bayesian framework that accounts for small samples, handles outliers, simulates high-fidelity data, and returns a posterior distribution over curve parameters that quantify estimation uncertainty. It uses a rectified-logistic function that estimates motor threshold and outperforms conventionally used sigmoidal alternatives in predictive performance, as demonstrated through cross-validation. In simulations, our method outperforms non-hierarchical models by reducing threshold estimation error on sparse data and requires fewer participants to detect shifts in threshold compared to frequentist testing. We present two common use cases involving electrical and electromagnetic stimulation data and provide an open-source library for Python, called hbMEP, for diverse applications.
}

\keywords{hierarchical Bayesian, hypothesis testing, motor threshold, transcranial magnetic stimulation, spinal cord stimulation, mathematical modeling, evoked potential}



\maketitle

%% file: sections/introduction.tex
\section{Introduction}\label{intro}
\begin{figure}[t]
    \centering
    \includegraphics[width=.6\textwidth]{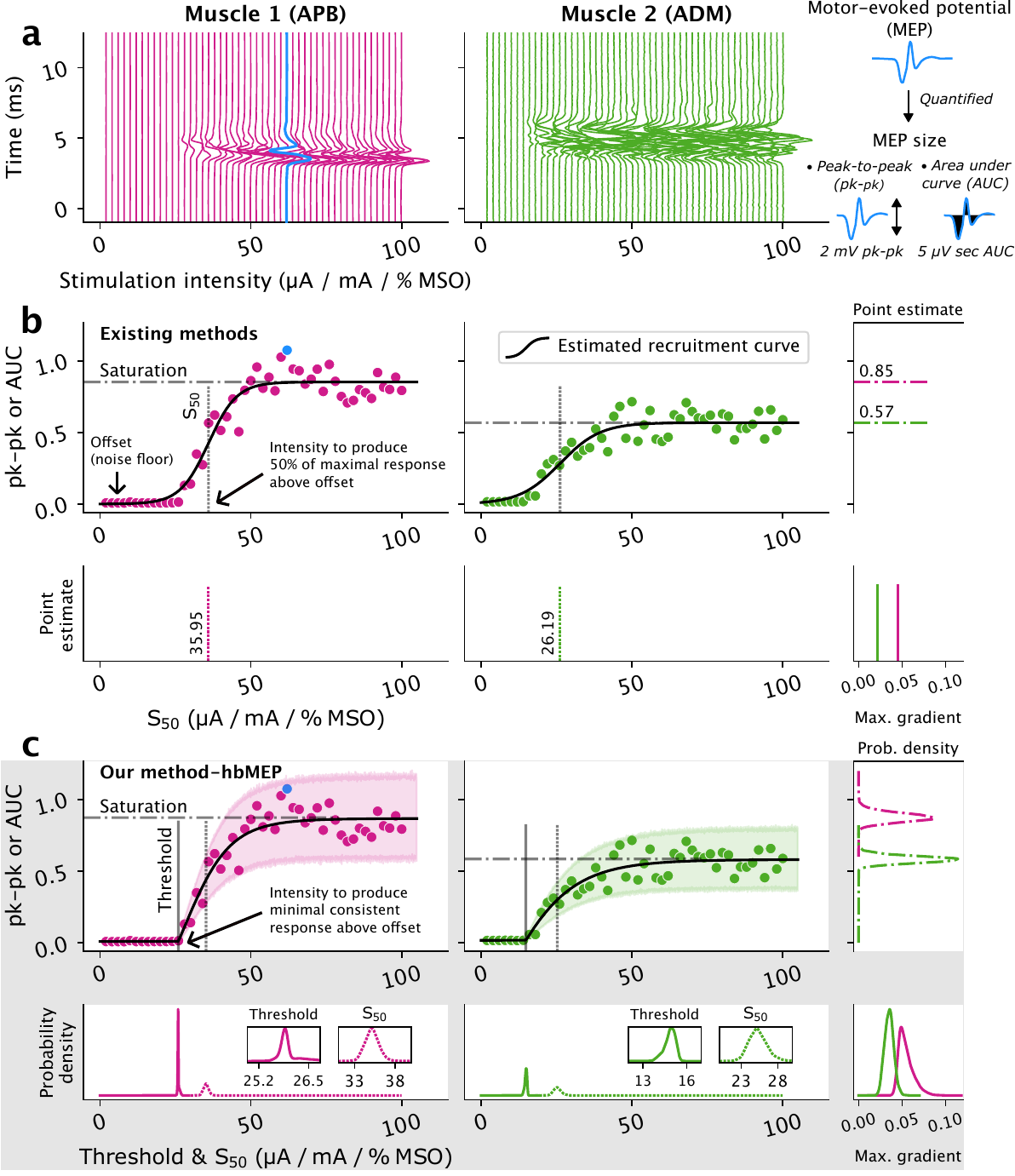}
    \caption{\textbf{Hierarchical Bayesian estimation of recruitment curves yields parameter estimates for each participant across multiple muscles simultaneously.}  \textbf{(a)} Example motor-evoked potentials (MEPs) recorded at different stimulation intensities from APB (left, magenta) and ADM (middle, green) muscles. The abscissa represents stimulation intensity which may be specified in units of current, e.g., $\mu$A or mA for spinal cord stimulation, or \% maximum stimulator output (\% MSO) for transcranial magnetic stimulation. Right panel: schematic quantification of MEPs into MEP size using either peak-to-peak (pk-pk) amplitude or area under the curve (AUC). \textbf{(b)} Example recruitment curves modeled as a 4-parameter Boltzmann sigmoid using least-squares minimization with the Nelder-Mead method. It provides only point estimates for curve parameters, lacks threshold estimate, and fails to capture sharp deflection from the offset. Bottom panels: point estimate of S$_{50}$. Top right: saturation. Bottom right: maximum gradient. \textbf{(c)} Example recruitment curves modeled as a 5-parameter rectified-logistic function within a hierarchical Bayesian framework. Shading represents the 95\% highest density interval (HDI) of the posterior predictive distribution. It accurately estimates the threshold, S$_{50}$, and saturation. For each parameter, estimation uncertainty can be quantified using the width of 95\% HDI. Data from multiple participants and muscles is handled simultaneously. Bottom panels: posterior distribution of threshold and S$_{50}$. Inset: zoom over posterior. Top right: saturation. Bottom right: maximum gradient.}\label{fig-introduction}
\end{figure}
Electrical and electromagnetic stimulation techniques, such as transcranial magnetic stimulation (TMS), spinal cord stimulation (SCS), and peripheral nerve stimulation, have become established methods for assessing nervous system function. This includes monitoring and planning clinical interventions or mapping muscle activation \citep{macdonald_intraoperative_2013, picht_preoperative_2011, sayenko_spinal_2015, hofstoetter_spinal_2021, greiner_recruitment_2021, mcintosh_intraoperative_2023}, measuring the extent of injury and tracking recovery \citep{chen_clinical_2008,millet_electrical_2011,balbinot_segmental_2023}, and evaluating the efficacy of therapeutic interventions, including neuromodulation \citep{stefan_induction_2000,bunday_motor_2012,pal_spinal_2022}.
When targeted to the sensorimotor system, these techniques produce motor-evoked potentials (MEPs) in multiple muscles (Fig. \ref{fig-introduction}a), which can be quantified by MEP size (Fig. \ref{fig-introduction}b,c), typically measured in terms of peak-to-peak (pk-pk) voltage or area under the curve (AUC). A key aspect of these techniques is the recruitment curve, which characterizes MEP size growth as a non-linear, non-decreasing function of stimulation intensity.

Accurate estimation of recruitment curves is critically important to assess nervous system state, including corticospinal excitability, and evaluate therapeutic efficacy \citep{devanne_input-output_1997,ridding_stimulusresponse_1997,boroojerdi_mechanisms_2001,rosenkranz_differential_2007,houdayer_effects_2008,moller_hysteresis_2009,nardone_assessment_2015,koponen_transcranial_2024}. The recruitment curve exhibits a characteristic S-shape, with a steep increase above the threshold and a plateau phase at high intensities. Its core properties (see Fig. \ref{fig-introduction}b,c) include offset (background noise floor of recording), saturation (upper asymptotic or maximal MEP size), growth rate (how MEP size increases with increasing intensity), S$_{50}$ (intensity to produce 50\% of maximal response above offset), and threshold (intensity to produce minimal consistent response above offset). While each of these properties may have a specific neurophysiological interpretation, both the threshold and S$_{50}$ are crucial for inferring changes in corticospinal excitability \citep{devanne_input-output_1997,feil_brain_2010,farzan_single-pulse_2014,nardone_assessment_2015}.

Current approaches \citep{van_de_ruit_novel_2019,ratnadurai_giridharan_motometrics_2019,skelly_mep-art_2020,hassan_brain_2022-1} to estimate recruitment curve parameters predominantly use sigmoidal or S-shaped functions (Fig. \ref{fig-introduction}b, black line) and rely on numerical optimization methods applied to non-convex search spaces, which are susceptible to suboptimal solutions and provide only point estimates. These techniques typically require a large number of samples to recover meaningful parameters with high accuracy. However, collecting an adequate number of samples is often infeasible due to constraints such as experimental time \citep{mcintosh_intraoperative_2023}, discomfort to participants \citep{manson_relationship_2020}, and the risk of inadvertent neuromodulation when large numbers of stimuli are delivered \citep{antal_no_2004,hassanzahraee_longer_2019}.

Moreover, the conventional approach of modeling recruitment curves using sigmoidal functions \citep{devanne_input-output_1997,pitcher_age_2003,klimstra_sigmoid_2008,kukke_efficient_2014,smith_locomotor_2015,murray_transspinal_2019,de_freitas_selectivity_2021} primarily aims to estimate the S$_{50}$ parameter, which is subsequently used to test hypotheses related to shifts in this parameter. By definition, estimation of S$_{50}$ is contingent upon observing adequate saturation in data, a condition rarely met due to the discomfort experienced by participants at higher stimulation intensities \citep{kukke_efficient_2014}. Conversely, the threshold can be estimated accurately independent of saturation, making it a more reliable parameter for testing. However, sigmoidal functions cannot be used to estimate the threshold \citep{devanne_input-output_1997,kukke_efficient_2014}. Previously, thresholds were estimated using a rectified-linear function \citep{willer_hypoxia_1987,devanne_input-output_1997,malone_closed-loop_2022-1,mcintosh_intraoperative_2023}, which proved overly simplistic as it does not capture the curvature in data.

In contrast, Bayesian methodology has shown great potential for improving the statistical modeling process \citep{gelman_bayesian_1995}. We introduce a hierarchical Bayesian framework that accounts for small samples, handles outliers, and returns a posterior distribution (Fig. \ref{fig-introduction}c, bottom and side panels) over curve parameters that quantify estimation uncertainty. Our method uses a rectified-logistic (Fig. \ref{fig-introduction}c, black line) function that estimates both the threshold and S$_{50}$. Through cross-validation on empirically obtained TMS and SCS data, we show that the rectified-logistic function outperforms traditionally used sigmoidal alternatives in predictive performance. The generative hierarchical Bayesian framework enables simulation of high-fidelity synthetic data for model comparison and optimizing experimental design. In simulations, we illustrate the robustness and efficiency of our framework, which outperforms conventionally used non-hierarchical models by minimizing threshold estimation error on sparse data. Additionally, we show that Bayesian estimation requires fewer participants to achieve equivalent statistical power and produces fewer false positives compared to frequentist null hypothesis testing when detecting shifts in threshold. Finally, we present two common use cases involving TMS and SCS data and introduce a library for Python, called hbMEP, for diverse applications.

%% file: sections/results.tex
\section{Results}\label{results}
\subsection{Accurate threshold estimation on sparse data}\label{res-accuracy}
\begin{figure}[t]
    \centering
    \includegraphics[width=\textwidth]{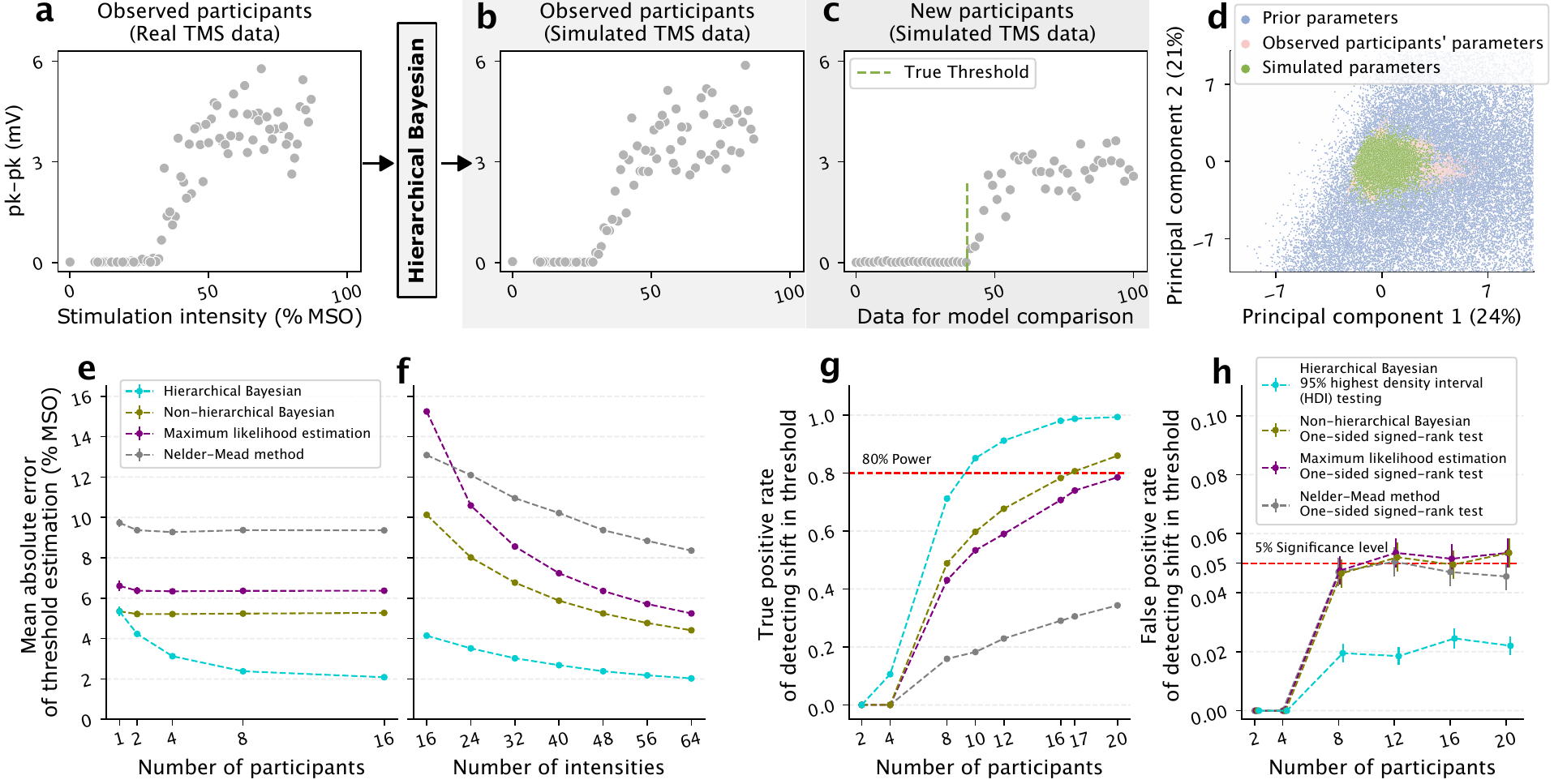}
    \caption{
        \textbf{(a--d) Generative hierarchical Bayesian model simulates high-fidelity synthetic data.} \textbf{(a)} Example participant from human TMS data that is used by hierarchical Bayesian model to estimate the participant- and population-level parameters. \textbf{(b)} Data simulated from the model conditioned on estimated participant-level parameters. The model can replicate observed participants. \textbf{(c)} Data simulated from the model conditioned on estimated population-level parameters for subsequent model comparison. \textbf{(d)} Principal component analysis shows a large overlap between the new simulated parameters (green dots) and those estimated from observed TMS data (pink dots). Blue dots represent parameters simulated from the weakly informative prior predictive distribution. \textbf{(e--f) Hierarchical Bayesian structure improves threshold estimation accuracy over non-Bayesian and non-hierarchical models on simulated data.} \textbf{(e)} Hierarchical Bayesian model (HB) benefits from partial pooling across participants and uniquely reduces mean absolute error of threshold estimation as the number of participants increase. Error bars represent standard error of the mean. \textbf{(f)} For eight simulated participants, the HB model outperforms other non-hierarchical methods at all tested number of stimulation intensities and its advantage is most pronounced for low number of intensities, i.e., on sparse data. \textbf{(g--h) Bayesian estimation is more powerful and produces fewer false positives when detecting shifts in threshold compared to frequentist testing.} \textbf{(g)} Bayesian estimation requires fewer participants to achieve 80\% power when detecting a negative shift in threshold from pre- to post-intervention phase. Threshold differences were simulated from $\text{Normal}\left(\mu=-5, \sigma=2.5\right)$, where the null hypothesis (no shift) is false. \textbf{(h)} Comparison of false positive rate of Bayesian estimation against the set significance level of 5\% of signed-rank test. Bayesian estimation is more conservative in falsely rejecting the null hypothesis and maintains a false positive rate between 0 and 2.5\% (at $N=16$ mean $\pm$ sem: $2.45 \pm 0.35\%$, at $N=20$: $2.2 \pm 0.33\%$). The differences were simulated from $\text{Normal}\left(0, 2.5\right)$, where the null hypothesis is true. Except for (f), all simulations consisted of 48 equispaced intensities between 0--100\% maximum stimulator output (\% MSO).
        }\label{fig-accuracy-power}
\end{figure}
We integrated the rectified-logistic function into a hierarchical Bayesian framework (Methods \ref{meth-simulation}) and used the resultant model to simulate synthetic data that closely matched real TMS data (Fig. \ref{fig-accuracy-power}a--d). The model estimated both participant- and population-level parameters from TMS data, which consisted of 27 participants (Fig. \ref{fig-accuracy-power}a). Using the estimated participant-level parameters, the model successfully replicated observations from existing TMS participants (Fig. \ref{fig-accuracy-power}b). Additionally, the model was conditioned on the estimated population-level parameters to simulate new participants (Fig. \ref{fig-accuracy-power}c). Principal component analysis (Fig. \ref{fig-accuracy-power}d) showed a large overlap between the parameters of new simulated participants and those estimated from existing TMS participants, validating the quality of the synthetic data.

The generated synthetic data included ground truth values for curve parameters (Fig. \ref{fig-accuracy-power}c, green line), enabling a comparative analysis of different estimation methods based on their accuracy of recovering true parameters. We evaluated the hierarchical Bayesian model (HB) for its accuracy in threshold estimation and compared it against three non-hierarchical models: the conventionally used maximum likelihood estimation (ML) and Nelder-Mead (NM, minimizing residual sum of squares) methods, and an equivalent non-hierarchical Bayesian (nHB) model (see Methods \ref{meth-accuracy} for detailed implementation).

The hierarchical Bayesian model demonstrated improved accuracy in threshold estimation over non-Bayesian and non-hierarchical models (Fig. \ref{fig-accuracy-power}e,f). For a single participant, the mean absolute error ($e$) of the HB model is similar to nHB (Fig. \ref{fig-accuracy-power}a, $e_\text{nHB} - e_\text{HB}$ mean $\pm$ sem: $0.01\pm 0.02$), which is lower than non-Bayesian models ($e_\text{ML} - e_\text{HB}$ $1.26 \pm 0.09$, $e_\text{NM} - e_\text{HB}$ $4.39 \pm 0.22$). As the number of participants increase, the HB model further reduces error over non-hierarchical models (for $N=16$ participants, $e_\text{nHB} - e_\text{HB}$ $3.19\pm 0.05$, $e_\text{ML} - e_\text{HB}$ $4.29 \pm 0.06$, $e_\text{NM} - e_\text{HB}$ $7.27 \pm 0.05$). In contrast, the errors for non-hierarchical models remain constant irrespective of the number of participants ($N$) used for analysis ($e_{N=1} - e_{N=16}$ for nHB $0.08 \pm 0.23$, ML $0.24 \pm 0.25$, NM $0.37 \pm 0.19$). The HB model also accounts for small sample sizes, where its advantage is most apparent with low number of intensities (Fig. \ref{fig-accuracy-power}f). For instance, with only 16 samples, the error difference between the nHB and HB models ($e_\text{nHB} - e_\text{HB}$) is $5.98 \pm 0.08$, with a 59\% reduction in error over the nHB model. In contrast, for a relatively large number of 64 samples, the difference is less pronounced at $2.38 \pm 0.06$, with a reduction of 54\%.

\subsection{Bayesian estimation for detecting shift in threshold}\label{res-power}
The flexibility of a hierarchical Bayesian model allows for modeling differences in participant-level thresholds, for example, between pre- and post-intervention phases (Methods \ref{meth-power}). These differences can be summarized across multiple participants using a population-level parameter, with its 95\% highest density interval (HDI) used for hypothesis testing \citep{gelman_type_2000,gelman_why_2009,kruschke_bayesian_2013,kruschke_doing_2014,vincent_hierarchical_2016}. To evaluate statistical power in the context of assessing effectiveness of an intervention, we simulated two hypothetical scenarios: one with a negative shift in the threshold from pre- to post-intervention, and another with no shift. A negative shift indicates that the intervention results in a lower threshold, thereby facilitating easier muscle activation. For Bayesian estimation, the decision rule rejects the null hypothesis (no shift) if the 95\% HDI is entirely left of zero. This was compared with a one-sided Wilcoxon signed-rank test on point estimates of pairwise threshold differences estimated using non-hierarchical models (see Methods \ref{meth-power} for detailed implementation of each model; a $t$-test was not applicable due to non-normality of estimated differences indicated by Shapiro-Wilk test). The significance level was set at 5\%, with the alternative hypothesis (negative shift) accepted if the $p$-value was less than 0.05.

Bayesian estimation required fewer participants to achieve 80\% power (same as true positive rate or TPR, given null hypothesis is false) compared to the non-hierarchical models (Fig. \ref{fig-accuracy-power}g). The nHB model required 17 participants to achieve 80\% power (at $N=17$ participants, TPR mean $\pm$ sem: $80.65 \pm 0.88\%$), while the HB model required only 10 participants (at $N=10$, TPR $85.15 \pm 0.79\%$), which is a reduction of 41\% in the number of participants. Additionally, the false positive rate (FPR) of Bayesian estimation remained well below the significance level of 5\% (Fig. \ref{fig-accuracy-power}h). It is more conservative in falsely rejecting the null hypothesis, in the sense that the FPR based on its 95\% HDI stays between 0 and 2.5\% (at $N=16$ FPR $2.45 \pm 0.35\%$, and at $N=20$ FPR $2.2 \pm 0.33\%$ when TPR $99.35 \pm 0.18\%$ is close to one), a known effect in hierarchical Bayesian models \citep{gelman_type_2000}.

\subsection{Choice of recruitment curve function}\label{res-choice}
\begin{figure}[t]
    \centering
    \includegraphics[width=\textwidth]{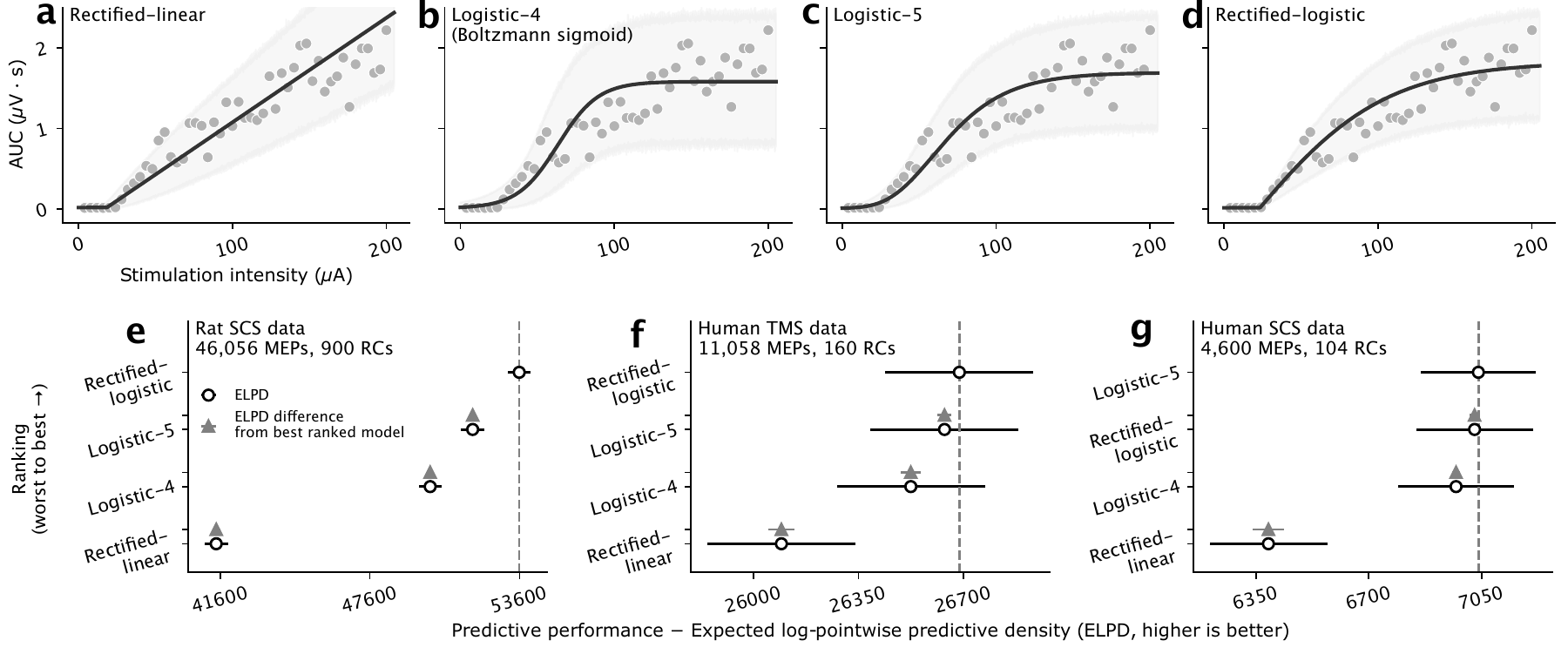}
    \caption{\textbf{Rectified-logistic function has at par or superior predictive performance compared to traditional alternatives while having the unique advantage of estimating threshold along with curvature and saturation.} \textbf{(a)} Example recruitment curve fitted to rat epidural SCS data using 3-parameter rectified-linear function. It underestimates the threshold at low intensities due to curvature in data, and subsequently overshoots at higher intensities while failing to capture saturation. Gray dots represent MEP size data, black line shows the curve, and gray shading represents the 95\% HDI of the posterior predictive distribution. \textbf{(b)} 4-parameter logistic-4 (Boltzmann sigmoid) is symmetric about its inflection point and saturates early while underestimating the sharp deflection from offset. \textbf{(c)} 5-parameter logistic-5 has improved deflection and saturation. \textbf{(d)} 5-parameter rectified-logistic function is flexible enough to accurately capture the deflection, curvature, and saturation, resulting in narrower 95\% HDI. \textbf{(e)} Predictive performance measured with expected log-pointwise predictive density (ELPD) leave-one-out cross-validation on rat SCS dataset. Black circles represent mean ELPD score, black bars are standard error of mean ELPD, gray triangles are mean pairwise ELPD difference from the best-ranked rectified-logistic model ($\Delta_\mu$), and gray bars are standard error of the mean ELPD difference ($\Delta_\text{sem}$). \textbf{(f)} As for (e), but on human TMS dataset. \textbf{(g)} As for (e), but on human epidural SCS dataset. Rectified-logistic function outperforms the rectified-linear function and the most commonly used logistic-4 function (rat SCS $\Delta_\mu \pm \Delta_\text{sem}$: $3569.2 \pm 117.4$, human TMS $162 \pm 34.5$, human SCS $57.6 \pm 18.7$, $\Delta_\mu \geq 3\Delta_\text{sem}$) on all datasets. It outperforms logistic-5 on the largest tested rat SCS dataset ($1866.2 \pm 93.6,\;\Delta_\mu \geq 3\Delta_\text{sem}$), and has comparable performance on human TMS ($49.6\pm 22.9$) and SCS ($-11.8 \pm 15.8$) datasets.}\label{fig-cross_validation_functional}
\end{figure}
The various choices for modeling recruitment curves include a 3-parameter rectified-linear function \citep{willer_hypoxia_1987,devanne_input-output_1997,malone_closed-loop_2022-1,mcintosh_intraoperative_2023} (Fig. \ref{fig-cross_validation_functional}a) and the most commonly used 4-parameter logistic-4 (also, Boltzmann sigmoid) function \citep{devanne_input-output_1997,klimstra_sigmoid_2008,kukke_efficient_2014,smith_locomotor_2015,murray_transspinal_2019,de_freitas_selectivity_2021} (Fig. \ref{intro}b, \ref{fig-cross_validation_functional}b). Additionally, a 5-parameter logistic-5 function \citep{pitcher_age_2003} (Fig. \ref{fig-cross_validation_functional}c) is a more generalized version of logistic-4 that is not necessarily symmetrical about its inflection point.

We evaluated these recruitment curve functions and the rectified-logistic function for their out-of-sample predictive performance using approximate leave-one-out cross-validation (PSIS-LOO-CV) \citep{vehtari_practical_2017} (Methods \ref{meth-choice}) on empirically obtained human and rat datasets for spinal cord and transcranial magnetic stimulation. The rectified-logistic function demonstrated superior predictive accuracy over the logistic-4 function on all datasets (Fig. \ref{fig-cross_validation_functional}e--g, rat SCS $\Delta_\mu \pm \Delta_\text{sem} : 3569.2 \pm 117.4$, human TMS $162 \pm 34.5$, human SCS $57.6 \pm 18.7$, $\Delta_\mu \geq 3\Delta_\text{sem}$ on all datasets). It also outperformed logistic-5 on the largest tested rat SCS dataset (Fig. \ref{fig-cross_validation_functional}e, $1866.2 \pm 93.6$, $\Delta_\mu \geq 3\Delta_\text{sem}$). For human TMS (Fig. \ref{fig-cross_validation_functional}f, $49.6\pm 22.9$) and SCS (Fig. \ref{fig-cross_validation_functional}g, $-11.8 \pm 15.8$) datasets, it maintained comparable  performance to logistic-5.

While the logistic functions are standard for estimating S$_{50}$, they lack an explicit parameter for the threshold, which can neither be derived from their equations since they are smooth functions. Conversely, the rectified-linear function includes a threshold parameter but exhibits suboptimal predictive performance (Fig. \ref{fig-cross_validation_functional}e--g). Estimation of S$_{50}$ requires observing adequate amount of saturation in data, whereas the threshold can be estimated accurately independent of that (Supplementary Fig. \ref{sufig-saturation}). Given that observing saturation is rare due to discomfort at higher stimulation intensities, this makes the threshold a more reliable parameter for inferring changes in corticospinal excitability. Therefore, the rectified-logistic function addresses the limitations of traditional logistic functions by enabling estimation of threshold while either surpassing or matching their predictive performance on all datasets.

\subsection{Robustness to outliers}\label{res-robustness}
\begin{figure}[t]
    \centering
    \includegraphics[width=\textwidth]{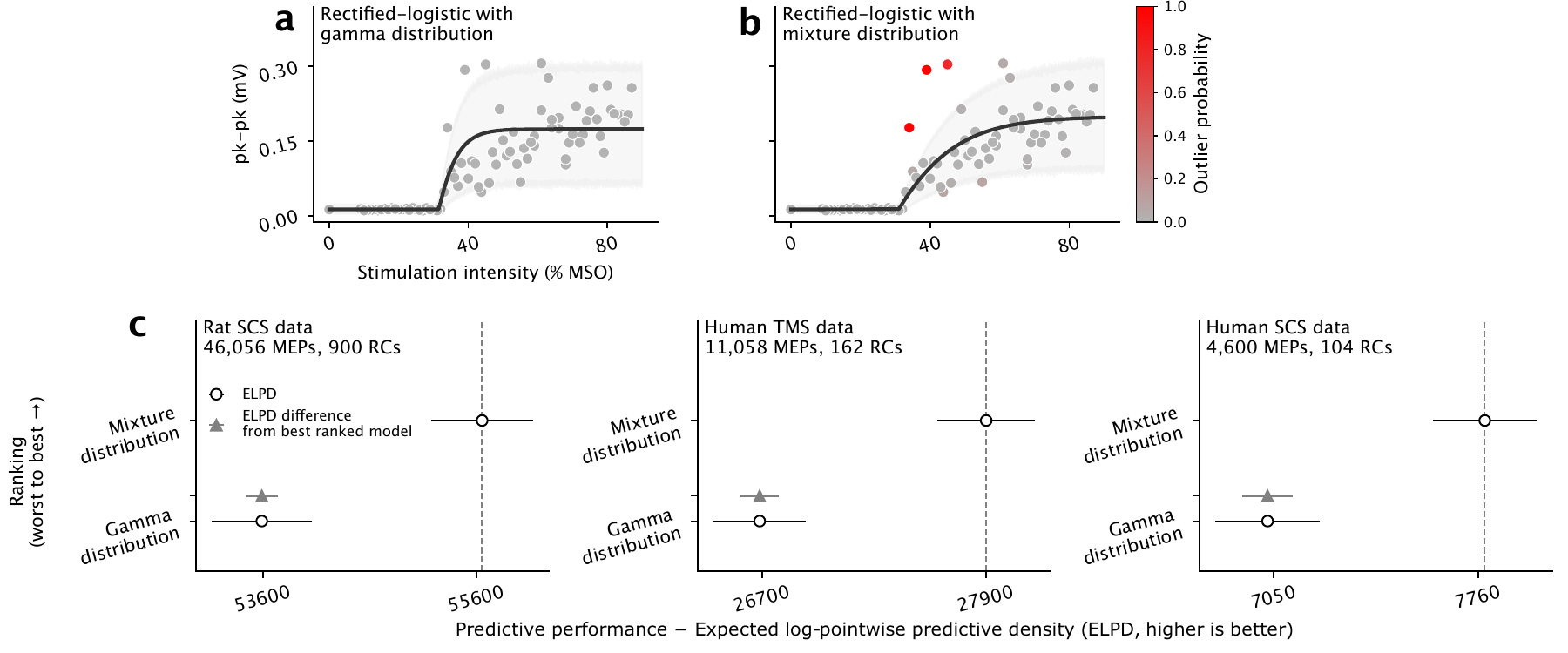}
    \caption{\textbf{Mixture model accounts for outliers and further improves predictive performance.} \textbf{(a)} Example recruitment curve fitted to human TMS data using rectified-logistic function within Gamma observation model. It overestimates the growth rate and saturates early due to presence of outliers. \textbf{(b)} Mixture extension of Gamma observation model is robust to outliers, resulting in narrower 95\% HDI of the posterior predictive distribution. Additionally, it returns an outlier probability for each observed sample which enables automatic outlier classification. Dots are colored by outlier probabilities. \textbf{(c)} Predictive performance measured with expected log-pointwise predictive density (ELPD) leave-one-out cross-validation. Gray triangles represent the mean pairwise ELPD difference from the best-ranked mixture model ($\Delta_\mu$), and gray bars are standard error of mean ELPD difference ($\Delta_\text{sem}$). The predictive performance of mixture model is significantly better $\left(\Delta_\mu \geq 3\Delta_\text{sem}\right)$ on rat SCS ($\Delta_\mu \pm \Delta_\text{sem} : 2061 \pm 150.5$), human TMS ($1214.5 \pm 103.7$), and human SCS ($754.1 \pm 88.8$) datasets.}\label{fig-cross_validation_mixture}
\end{figure}
Inaccuracies in model fitting often arise from sources of variability that occur independently of stimulation intensity, such as fasciculations, movement, or technical anomalies. To account for these rare occurrences, we introduced a mixture extension of our Gamma observation model (Methods \ref{meth-observation}, \ref{meth-mixture}). This extension assigns a small, learnable probability that an observed sample comes from a broad distribution independent of stimulation intensity.

This adjustment yields robust estimates that are otherwise biased by outliers (Fig. \ref{fig-cross_validation_mixture}a, overestimated growth rate) and enables automatic outlier classification by returning an outlier probability for each observed sample (Fig. \ref{fig-cross_validation_mixture}b, red dots). It further improved the predictive accuracy on all datasets (Fig. \ref{fig-cross_validation_mixture}c, rat SCS $2061 \pm 150.5$, human TMS $1214.5 \pm 103.7$, human SCS $754.1 \pm 88.8$, $\Delta_\mu \geq 3\Delta_\text{sem}$ on all datasets).

\subsection{Within-participant comparison}\label{res-within}
\begin{figure}[t]
    \centering
    \includegraphics[width=\textwidth]{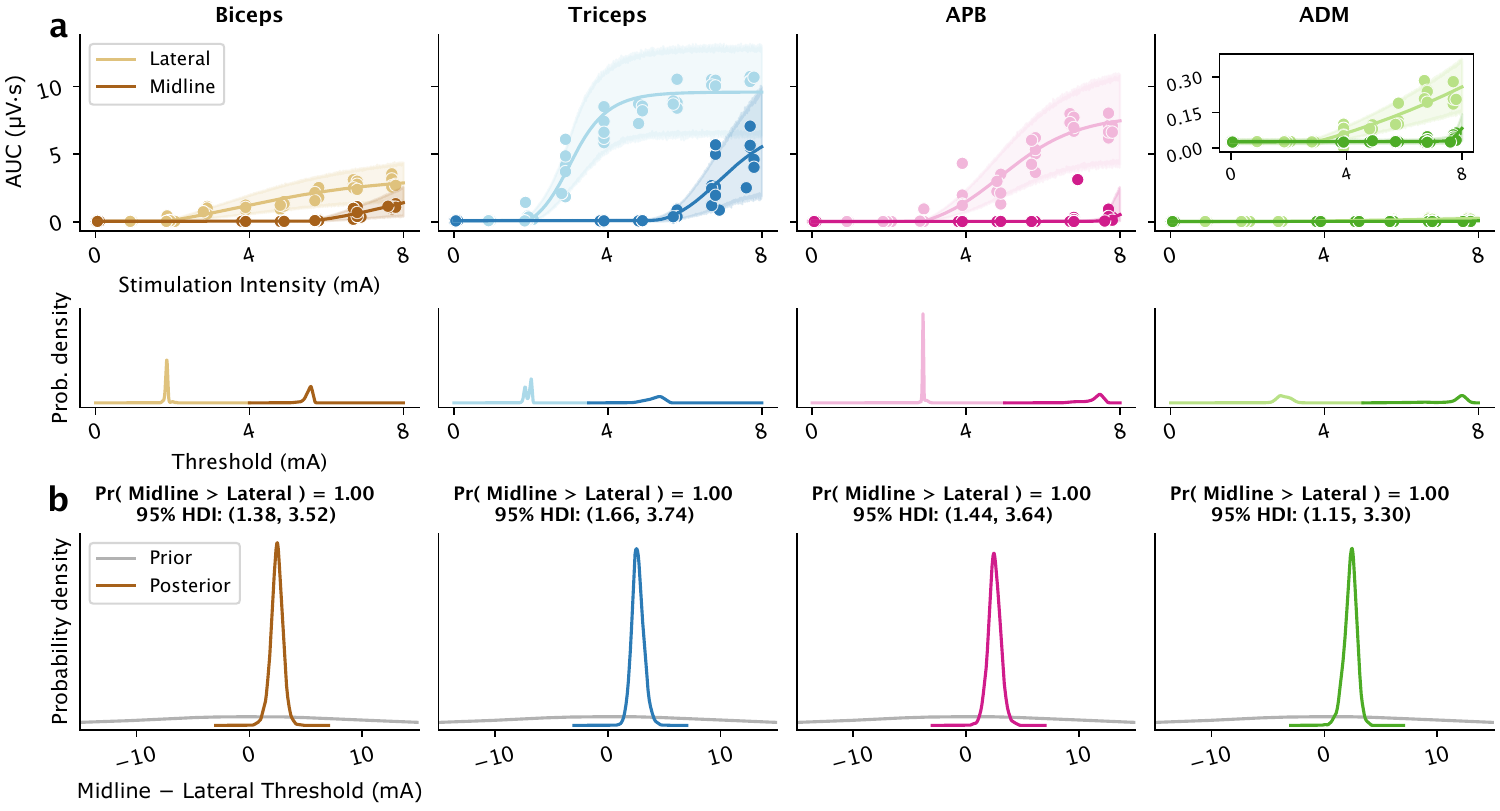}
    \caption{\textbf{Comparison of midline versus lateral stimulation thresholds on human epidural SCS data.} \textbf{(a)} Example participant showing lateral (light color) and midline (dark color) stimulation. Inset: zoom to show presence of threshold, despite small MEP size. Bottom panels: posterior distribution of threshold. \textbf{(b)} Posterior distribution of shift between midline and lateral thresholds summarized across 13 participants who underwent intraoperative surgery. A priori, the model assumes no shift, indicated by a flat prior (gray distribution) centered at zero. The 95\% HDI for biceps $(1.38, 3.52)$, triceps $(1.66, 3.74)$, APB $(1.44, 3.64)$, and ADM $(1.15, 3.30)$ are all right of zero, which indicates strong evidence that lateral stimulation resulted in significantly lower thresholds for arm and hand muscles.}\label{fig-use_cases_within}
\end{figure}
To validate our method's applicability to real data, we conducted a secondary analysis of epidural SCS data collected from 13 participants who underwent clinically indicated cervical spine surgery, which resulted in lower thresholds when stimulating laterally as compared to midline \citep{mcintosh_intraoperative_2023}. Utilizing our hierarchical Bayesian approach (Methods \ref{meth-use-within}), we compared the midline and lateral stimulation thresholds (Fig. \ref{fig-use_cases_within}) for arm and hand muscles. Note that in analyses such as these, we would like to know for each individual muscle whether or not a statistically significant effect is present. The 95\% HDI of threshold differences summarized across all participants lies entirely right of zero for all muscles (Fig \ref{fig-use_cases_within}b), suggesting strong evidence that lateral stimulation resulted in significantly lower thresholds, facilitating easier activation of arm and hand muscles.

\subsection{Between-groups comparison}\label{res-between}
\begin{figure}[t]
    \centering
    \includegraphics[width=\textwidth]{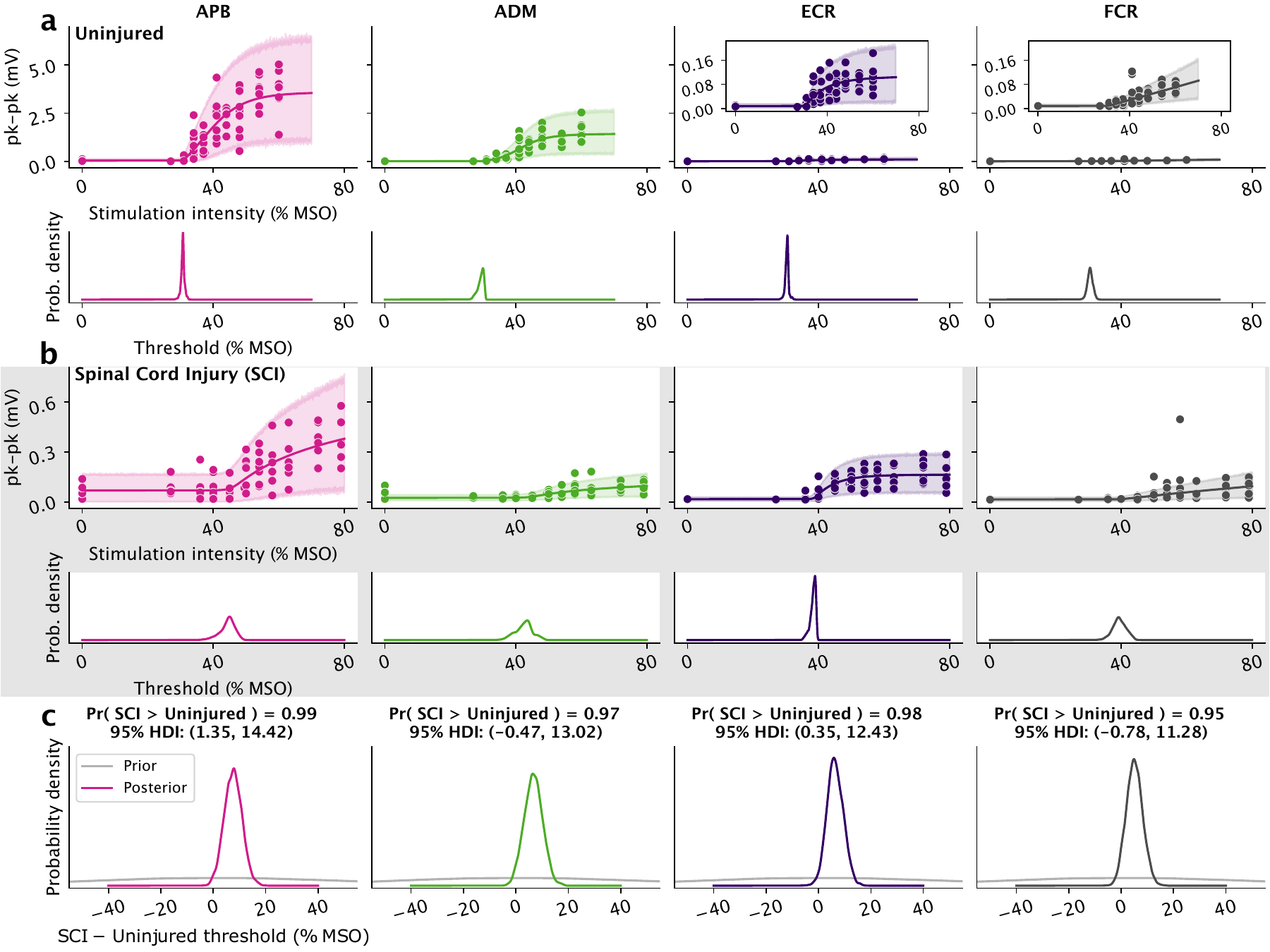}
    \caption{\textbf{Comparison of TMS thresholds in uninjured participants and participants with SCI.} \textbf{(a)} Example uninjured participant. Inset: zoom to show presence of threshold, despite small MEP size. Bottom panels: posterior distribution of threshold. \textbf{(b)} Example participant with spinal cord injury (SCI). \textbf{(c)} Posterior distribution of threshold difference summarized between 14 uninjured and 13 SCI participants. A priori, the model assumes no difference, indicated by a flat prior (gray distribution) centered at zero. The 95\% HDI for the target APB muscle $(1.35, 14.42)$, and ECR $(0.35, 12.43)$ are entirely right of zero. This, together with the Bayesian probabilities, indicates that spinal cord injury is associated with higher thresholds in hand and wrist muscles.}\label{fig-use_cases_between}
\end{figure}
We also used a hierarchical Bayesian model (Methods \ref{meth-use-group}) to compare thresholds between 14 uninjured participants and 13 participants with spinal cord injury (SCI). SCI participants spanned American Spinal Injury Association Impairment Scale (AIS) classifications A--D, with motor levels ranging from C2 to T1. Figure \ref{fig-use_cases_between} shows the result for hand and wrist muscles. The 95\% HDI interval that summarizes the differences in thresholds between the SCI and uninjured group of participants is entirely right of zero for the target APB muscle, and ECR, indicating that spinal cord injury is associated with significantly higher thresholds for these muscles. Although not significant, the Bayesian estimation method indicates with high probability (ADM: 97\%, FCR: 95\%) that a similar effect is present for the other muscles. This is a more satisfactory and informative conclusion than a frequentist test, which would only provide a dichotomous outcome of significant or not significant based on a $p$-value.

\subsection{Optimizing experimental design}\label{res-reps}
Researchers often need to determine the optimal number of repetitions for each stimulation intensity during experiments to accurately estimate thresholds. For instance, when administering 64 pulses in a session, a researcher might choose to test 64 equispaced intensities with a single repetition each, 16 equispaced intensities with four repetitions each, or 8 equispaced intensities with eight repetitions each, and so on.

Our generative framework enables the examination of such experimental strategies. For 8 participants of synthetic TMS data (Results \ref{res-accuracy}, Methods \ref{meth-use-optimization}), we generated up to 8 observations per stimulation intensity, for a total of 64 intensities equispaced between 0 to 100\% MSO. We used our model to estimate thresholds using one, four and eight observations per intensity. The results indicate that performing experiments without repetition, i.e., testing each intensity only once, provides the most efficient threshold estimation (Supplementary Fig. \ref{sufig-repetitions}). Notably, the largest improvement in accuracy is observed when the number of pulses are low. For instance, distributing 32 pulses evenly across the full intensity range reduced the mean absolute error from 6.18 $\pm$ 0.04\% (mean $\pm$ sem) with 8 repetition per 4 equispaced intensities to 3.02 $\pm$ 0.04\% with single repetition per 32 equispaced intensities.

%% file: sections/discussion.tex

\section{Discussion}\label{discussion}
Our method integrates a rectified-logistic function within a hierarchical Bayesian framework to estimate MEP size recruitment curves. We validated the applicability of our model on TMS and SCS data, and demonstrated its utility in analyzing the threshold parameter in two common scenarios--assessing shifts within participants undergoing repeat experiments and comparing thresholds between different participant groups. Its generative capability was verified through simulation of synthetic data that closely matched real observations. In simulations, our approach improved threshold estimation accuracy on sparse data and increased statistical power for detecting shifts in threshold. To streamline the estimation process, we developed an open-source Python library, called hbMEP, which enables users to model recruitment curves and test hypotheses using the advantages of hierarchical Bayesian estimation.

Using a recruitment curve function that does not appropriately represent the data can produce systematic errors in parameter estimates, thereby compromising subsequent analysis. We demonstrated how MEP size recruitment curves, whether derived from TMS or epidural SCS, can be modeled using the same 5-parameter rectified-logistic function. Through Bayesian cross-validation on TMS and SCS data, we determined that 3-parameter rectified-linear and 4-parameter logistic-4 functions are suboptimal for estimating recruitment curves. This is due to the strict assumptions these functions make--rectified-linear assumes linear growth post-threshold and logistic-4 assumes symmetry about its inflection point. We found that a more flexible 5-parameter logistic-5 performs significantly better, but it is limited to estimating the S$_{50}$ and not the threshold. However, we illustrated how the estimation of S$_{50}$ depends on the amount of saturation observed in data, whereas the threshold can be accurately estimated independent of that. To shift focus on analyzing a more reliable threshold parameter, we developed a 5-parameter rectified-logistic function that matches or exceeds the predictive performance of logistic-5 and has a direct representation of threshold as a parameter.

Our generative framework enables simulation of synthetic data that closely matches real data, providing a robust testbed for benchmarking estimation methods and optimizing experimental design. Electrophysiology experiments are often conducted under constraints involving time and participant burden, resulting in undersampled data. On simulated data, we demonstrated how hierarchical Bayesian estimation accounts for small samples and improves threshold estimation accuracy. This is enabled through partial pooling \citep{gelman_bayesian_1995,gelman_bayesian_2006} which allows it to leverage data across participants. The parametric estimation for each individual is informed and strengthened by the data from other individuals who share similar characteristics. In contrast, this sharing of information is absent in non-hierarchical models. We presented a model for detecting shifts in threshold and demonstrated how Bayesian estimation requires fewer participants to achieve equivalent statistical power while producing fewer false positives compared to frequentist testing. Additionally, this model can also be used to optimize stimulation parameters such as electrode orientation, size, and pulse shape, where multiple comparisons are needed. Hierarchical Bayesian models typically address issues with multiple comparisons by making comparisons appropriately more conservative through partial pooling \citep{gelman_bayesian_1995,gelman_type_2000,gelman_why_2009}.

Our parametric representation of the threshold differs in approach from the resting motor threshold traditionally used in TMS studies. The predominant definition of resting motor threshold is the minimum stimulation intensity that produces a predefined MEP size in at least 50\% of repetitions \citep{rossini_non-invasive_2015}. The standard definition uses 50 $\mu$V peak-to-peak amplitude as the predefined size for resting motor threshold and 200 $\mu$V peak-to-peak for active motor threshold \citep{rossini_non-invasive_2015}. The adoption of these criteria was partially influenced by the technological limitations of amplifiers and electromyography equipment available at the time, and its relevance stems more from widespread adoption than any inherent biological construct. Furthermore, while a 50 $\mu$V predefined MEP size may be suitable for human resting TMS experimental conditions, it is not clear how this definition generalizes across different species, stimulation modalities, or MEP size metrics such as AUC. In contrast, the rectified-logistic function provides a consistent parametric representation of the threshold as a deflection of MEP size from the estimated offset. Additionally, the resting motor threshold conflates a change in parametric threshold with the initial rate of increase of that activation. For example, a participant's muscle may begin to show clear and consistent MEPs at a given stimulus intensity, but if those MEPs are below the customary 50 $\mu$V cutoff, that intensity would still be defined as subthreshold, necessitating a higher stimulus intensity to reach the cutoff. This leads to situations in which subthreshold stimulation does not mean there is no evoked potential. This discrepancy can introduce undesirable complexity to certain experimental paradigms. Nevertheless, given the extensive use of the resting motor threshold in previous studies, we deemed it important to maintain backward compatibility. Our library for Python enables recovering the traditional motor threshold as a post-processing step by intersecting the estimated recruitment curves with a predefined MEP size, provided that this predefined size is between the estimated offset and saturation. For the same reason of compatibility, our library also enables estimating recruitment curves using logistic-5, logistic-4, and rectified-linear functions while maintaining advantages of hierarchical Bayesian estimation.

Multiple sources of variability impact the size of the MEP as measured by peak-to-peak or AUC, and these sources were not exhaustively modeled in our approach. For instance, when applying epidural \citep{sharma_vivo_2021} or transcutaneous \citep{wu_posteroanterior_2020} SCS, posterior roots are thought to be preferentially activated. However, if sufficient stimulation intensity is applied, efferent fibers may also be activated such that this activation is represented in the first stretch of the MEP. Without a careful choice of the time window for calculating the MEP size, the measured recruitment curve may actually represent a mixture of efferent and afferent recruitment curves, which would not be accounted for by any S-shaped curve. Other sources of variability may exist that cannot be addressed through preprocessing alone and may require extensions of the proposed models.

In conclusion, our approach accounts for small samples, improves parametric estimation accuracy over non-hierarchical models, and achieves greater statistical power compared to frequentist testing. By enhancing accuracy on sparse experimental data, this approach reduces participant burden by decreasing the necessary duration and the number of stimuli required to probe each individual's neuromuscular parameters, reducing the risk of inadvertent neuromodulation, while simultaneously increasing the number of muscles across which these insights are obtained.

%% file: sections/methods.tex
\section{Methods}\label{methods}
In sections \ref{meth-modeling} and \ref{meth-hierarchical}, we describe the theory of hierarchical Bayesian modeling in the context of MEP size recruitment curves. Section \ref{meth-modeling} begins by discussing the various functions that are currently used for estimating recruitment curves, including the 3-parameter rectified-linear, 4-parameter logistic-4, and 5-parameter logistic-5 functions. We then introduce a 5-parameter rectified-logistic function that combines aspects of these functions while allowing for more accurate threshold estimation. We detail the gamma observation model and, in section \ref{meth-hierarchical}, its integration into a hierarchical Bayesian framework.

In sections \ref{meth-robustness}--\ref{meth-uses}, we detail how the methods were used to arrive at the results presented in the manuscript. Section \ref{meth-robustness} discusses the robustness and efficiency of the proposed methods. We describe the use of synthetic data to validate the performance of the hierarchical Bayesian model. This section also covers the accuracy of threshold estimation on sparse data and Bayesian estimation for detecting shifts in threshold. In section \ref{meth-choice}, we compare different recruitment curve functions using cross-validation and validate the choice of the rectified-logistic function. We then discuss an extension of the model to handle outliers using a mixture model. In section \ref{meth-uses}, we describe common use cases for hierarchical Bayesian models, including within-participant and between-groups comparisons, and optimizing experimental design.

Section \ref{meth-reproducibility} mentions about the statistical methods used and the reproducibility of presented analyses. Sections \ref{meth-human_data} and \ref{meth-rat_data} describe the human TMS and SCS, and the rat SCS datasets, respectively.

\subsection{Modeling MEP size}\label{meth-modeling}
The various choices for modeling recruitment curves include a 3-parameter rectified-linear function \citep{willer_hypoxia_1987,devanne_input-output_1997,malone_closed-loop_2022-1,mcintosh_intraoperative_2023} (Eq. \ref{eq411}, Fig. \ref{fig-cross_validation_functional}a) and a 4-parameter logistic-4 function \citep{devanne_input-output_1997,klimstra_sigmoid_2008,kukke_efficient_2014,smith_locomotor_2015,murray_transspinal_2019,de_freitas_selectivity_2021} (Eq. \ref{eq412}, Fig. \ref{intro}b, \ref{fig-cross_validation_functional}b). Additionally, a 5-parameter logistic-5 \citep{pitcher_age_2003} (Eq. \ref{eq413}, Fig. \ref{fig-cross_validation_functional}c) is a more generalized version of logistic-4 and contains an extra parameter $v$ to control near which asymptote (lower $L$, or upper $L + H$) the maximum growth or inflection point occurs. In contrast to logistic-4, the logistic-5 is not necessarily symmetrical about its inflection point.
\begin{align}
    &\text{Rectified-linear} & \forall a, b, L > 0\ \;\;&x \mapsto L + \max\left\{0, b\left(x - a\right)\right\}  \label{eq411}\\
    &\text{Logistic-4} & \forall a, b, L, H > 0\ \;\;&x \mapsto L + \frac{H}{1 + e^{-b\left(x-a\right)}}  \label{eq412}\\
    &\text{Logistic-5} & \forall a, b, v, L, H > 0\ \;\;&x \mapsto L + \frac{H}{\left\{1 + \left(2^v - 1\right)e^{-b\left(x-a\right)}\right\}^{1/v}}  \label{eq413}
\end{align}

Note that in the rectified-linear function (Eq. \ref{eq411}), parameter $a$ models the threshold, and in logistic functions (Eq. \ref{eq412}, \ref{eq413}) it models the $\text{S}_{50}$. $L$ represents the offset MEP size, $\left(L + H\right)$ defines the saturation, and $b$ is the growth rate. The logistic functions do not have a parameter for threshold, which can neither be derived from their equations since they are smooth functions; and the rectified-linear function does not have a parameter for $\text{S}_{50}$ since it does not saturate.

\subsubsection{Observation model}\label{meth-observation}
We introduce a 5-parameter rectified-logistic function (Eq. \ref{eq414}, Fig. \ref{fig-introduction}c, \ref{fig-cross_validation_functional}d) that can estimate both the threshold and $\text{S}_{50}$. Supplementary Fig. \ref{sufig-meth_default}a-f shows the effect of varying its parameters. Parameters $L, H, b$ have similar interpretation as in the logistic-4 function, and $a$ models the threshold. Equation \ref{eq415} gives the $\text{S}_{50}$ of the rectified-logistic function. Similar to logistic-5, there is an additional parameter $\ell$ (Supplementary Fig. \ref{sufig-meth_default}e) that controls the location of inflection point, whether near the offset $L$ or saturation $\left(L + H\right)$.
\begin{align}
    \shortintertext{For $a, b, L, \ell, H > 0$, define $\mathcal{F}: \mathbb{R} \to \mathbb{R}^{+}$ as}
    \mathcal{F}\left(x\right) &= L + \max\left\{0, -\ell + \frac{H + \ell}{1 + \left(\frac{H}{\ell}\right)e^{-b\left(x-a\right)}} \right\} \label{eq414}\\
    {\text{S}_{50}}\left(\mathcal{F}\right)&= a - \frac1b\ln\left(\frac{\ell}{H + 2\ell}\right) \label{eq415}
\end{align}

We use a gamma observation model in shape-rate parametrization (Eq. \ref{eq416}--\ref{eq418}) to model the relationship between MEP size $\left(y\right)$ and stimulation intensity $\left(x\right)$. Note that we model the expected MEP size (Eq. \ref{eq417}) as a rectified-logistic function of intensity, since $\mathbb{E}\left(y \mid x, \Omega, c_1, c_2\right) = \mu = \mathcal{F}\left(x\mid\Omega\right)$.
\begin{align}
    \shortintertext{For $c_1, c_2 > 0$}
    y \mid x, \Omega, c_1, c_2 &\sim \text{Gamma}\left(\mu\cdot\beta, \beta\right) \label{eq416}\\
    \mu & = \mathcal{F}\left(x \mid \Omega\right) \label{eq417} \hspace{40pt}{\Omega} = \left\{a, b, L, \ell, H\right\}\\
    \beta &= \frac{1}{c_1} + \frac{1}{c_2\cdot\mu} \label{eq418}
\end{align}

We chose a gamma distribution to capture the long-tailed distribution of MEP size around the recruitment curve. In Eq. \ref{eq418} we specify the rate parameter $\left(\beta\right)$ as a linear function of the reciprocal of expected MEP size $\left(\frac1\mu\right)$ with positive weights $\left(\frac1{c_1}, \frac1{c_2}\right)$ to capture the heteroskedastic spread that increases with increasing MEP size.

\subsubsection{Recruitment curves}
More generally, $\mathcal{F}$ (Eq. \ref{eq417}) is called the activation function or the recruitment curve in the context of modeling MEP size, which transforms the input stimulation intensity $\left(x\right)$, and links it to the expected MEP size $\mathbb{E}\left(y\mid x\right)$. $\mathcal{F}$ can be replaced by other available choices, including rectified-linear, logistic-4, or logistic-5.

\subsection{Hierarchical Bayesian Model}\label{meth-hierarchical}
\subsubsection{Default model}\label{meth-default}
The simplest form of a standard 3-stage hierarchical Bayesian model (Eq. \ref{eq421}--\ref{eq423}) in the context of modeling MEP size can be described as follows. Let there be $N_P \times N_M$ exchangeable sequences $\left\{\left({x_i}^{p}, {y_i}^{p,m}\right)_{i=1}^{n(p)} \mid p = 1\ldots N_P, m = 1\ldots N_M\right\}$ of MEP size ${y_i}^{p,m} \in \mathbb{R}^+$ recorded at stimulation intensity ${x_i}^p \in \mathbb{R}^+\cup\{0\}$ from muscle $m$ of participant $p$, for a total of $N_M$ muscles of $N_P$ participants. Here $n(p)$ denotes the number of intensities tested for participant $p$, which is independent of muscle $m$ since MEP size ${y_i}^{p,m}$ is recorded simultaneously from all muscles $m=1,\ldots N_M$ at a given intensity ${x_i}^p$.

The first stage of hierarchy is the participant-level (Eq. \ref{eq421}). It specifies the parametric model $P\left({y_i}^{p,m} \mid {x_i}^p, \theta^{p,m}\right)$ for each of the $N_M$ muscles of $N_P$ participants, and models the MEP size ${y_i}^{p,m}$ as a function of intensity ${x_i}^p$ and participant-level parameters $\theta^{p,m}$. In the second stage (Eq. \ref{eq422}), the participant-level parameters $\theta^{p,m}$ are assumed to be generated from a common distribution $P\left(\theta^{p,m} \mid \gamma\right)$ with population-level hyper-parameters $\gamma$. In the third stage (Eq. \ref{eq423}), the population-level hyper-parameters $\gamma$ are assumed to be unknown and assigned a weakly informative prior $P\left(\gamma\right)$, also called the hyperprior.
\begin{align}
    &\text{Stage I} &{y_i}^{p,m} &\sim P\left({y_i}^{p,m} \mid {x_i}^p, \theta^{p,m}\right)& \label{eq421}\\
    &\text{Stage II} &\theta^{p,m} &\sim P\left(\theta^{p,m} \mid \gamma\right)& \label{eq422}\\
    &\text{Stage III} &\gamma &\sim P\left(\gamma\right)& \label{eq423}
\end{align}
Supplementary Fig. \ref{sufig-meth_default}g,h specifies the default rectified-logistic model for human TMS data that was compared with other available recruitment curve functions in Results \ref{res-choice}.

\subsubsection{Within-participant comparison}\label{meth-within}
This section presents a hierarchical model that is useful for modeling shifts in curve parameters. This is applicable in settings where the same set of participants are tested for multiple experimental conditions (repeat measurements), such as pre- and post-intervention phases, stimulation locations (e.g., midline or lateral), or stimulation parameters (e.g., electrode size, stimulation frequency).

Supplementary Fig. \ref{sufig-meth_within} gives the graphical representation of such a model used to summarize differences in the threshold parameter. Here, we have the threshold $a^{p,c,m}$ of participant $p$, at tested condition $c$ and muscle $m$ given by,
\begin{align}
    a^{p, c, m} = \begin{cases}
        {{a}_\text{fixed}}^{p, m} & c = 1 \\
        {{a}_\text{fixed}}^{p, m} + {{a}_\Delta}^{p, c, m} & c > 1
    \end{cases}\label{eq424}
\end{align}

The threshold is broken (Eq. \ref{eq424}) into a fixed component $\left(c = 1\right)$ and a shift component $\left(\forall c > 1\right)$ that measures the difference from the fixed component. This shift component is parametrized by condition-level location $\left({\mu_{a_{\Delta}}}^{c,m}\right)$ and population-level scale $\left(\sigma_{a_\Delta}\right)$ hyperparameters. The location hyperparameter summarizes the shift of each tested condition $\left(\forall c > 1\right)$ from the fixed component $\left(c = 1\right)$ for each muscle. The scale parameter measures the overall variability in the estimated shifts.

Additionally, the location hyperparameter $\left({\mu_{a_{\Delta}}}^{c,m}\right)$ is partially pooled across conditions and muscles and it is given location $\left(\mu_{\mu_{a_\Delta}}\right)$ and scale $\left(\sigma_{\mu_{a_\Delta}}\right)$ hyperparameters. This partial pooling is done to account for multiple comparisons \citep{gelman_type_2000,gelman_why_2009}. A priori we assume there is no shift from the fixed component and $\mu_{\mu_{a_\Delta}}$ is given a flat prior symmetric about zero. Once the model is fit, the 95\% HDI of ${\mu_{a_{\Delta}}}^{c,m}$ posterior is used to assess the strength of the shift for condition $c$ and muscle $m$.

\subsubsection{Between-groups comparison}\label{meth-between}
Supplementary Fig. \ref{sufig-meth_between} gives a hierarchical model used to compare threshold between different groups of participants. The parameter of interest, threshold in this case, is parameterized by group-level location $\left({\mu_a}^{g,m}\right)$ and population-level scale $\left(\sigma_a\right)$ hyperparameters. The location parameter summarizes the thresholds of all participants belonging to a group. The scale parameter summarizes the overall variability in the estimated thresholds.

Additionally, this location hyperparameter $\left({\mu_a}^{g,m}\right)$ is partially pooled across groups and muscles and given location $\left(\mu_{\mu_a}\right)$ and scale $\left(\sigma_{\mu_a}\right)$ hyperparameters. Once the model is fit, the 95\% HDI of posterior difference ${\mu_{a}}^{g_{g_1},m} - {\mu_{a}}^{g_{g_2},m}$ is used to compare the parameter between groups $g_1$ and $g_2$ at muscle $m$.

\subsubsection{Extension to mixture model}\label{meth-mixture}
The models discussed so far can be extended to handle outliers by replacing the gamma distribution (Eq. \ref{eq416}) with a 2-component mixture of the gamma and a half-normal distribution. The resultant observation model is given as,
\begin{align}
    y \mid x &\sim \left(1 - q_y\right)\cdot\text{Gamma}\left(\mu\cdot\beta, \beta\right) + q_y \cdot \text{HalfNormal}\left(\sigma_{\text{outlier}}\right) \label{eq425} \\
    q_y &\sim \text{Bernoulli}\left(p_{\text{outlier}}\right) \label{eq426} \\
    p_{\text{outlier}} &\sim \text{Uniform}\left(0, C_{p_{\text{outlier}}}\right) \label{eq427} \\
    \sigma_{\text{outlier}} &\sim \text{HalfNormal}\left(C_{\sigma_{\text{outlier}}}\right)\label{eq428}
\end{align}

where $C_{p_{\text{outlier}}}, C_{\sigma_{\text{outlier}}} > 0$ are constants and $C_{p_{\text{outlier}}} < 1$ is chosen to be small, usually in the range $0.01 - 0.05$. Intuitively, this means that we expect roughly $1\% - 5\%$ of outliers to be captured by the half-normal distribution. The choice of $C_{\sigma_{\text{outlier}}}$ is based on the expected range of outliers. Supplementary Fig. \ref{sufig-meth_mixture} shows the mixture extension of the default model (Methods \ref{meth-default}).

\subsection{Robustness \& efficiency}\label{meth-robustness}
\subsubsection{Synthetic data and posterior predictive checks}\label{meth-simulation}
\noindent In Results \ref{res-accuracy} (Fig. \ref{fig-accuracy-power}a--d), we used the default model (Methods \ref{meth-default}, Supplementary Fig. \ref{sufig-meth_default}g,h) to estimate participant-level and population-level parameters from TMS data. We used data from the APB muscle ($N_M = 1$), which was the target muscle for 21 of the total 27 participants ($N_P = 27$). The model was conditioned on estimated participant-level parameters $\left(c_1, c_2, a \ldots H\right)$ to replicate the observed participants (Fig. \ref{fig-accuracy-power}b). It was conditioned on estimated population-level parameters $\left(\sigma_{c_1}, \sigma_{c_2}, \mu_a \ldots \sigma_H\right)$ to simulate new participants (Fig. \ref{fig-accuracy-power}c).

Principal component analysis (PCA) was used to visualize the participant-level parameters on the cartesian plane (Fig. \ref{fig-accuracy-power}d). The PCA map was fit on parameters estimated from existing TMS participants (Fig. \ref{fig-accuracy-power}d, pink dots). The map was used to project parameters simulated from the prior predictive distribution (blue dots) and parameters of the new simulated participants (green dots) on the cartesian plane.

\subsubsection{Accurate threshold estimation on sparse data}\label{meth-accuracy}
\noindent The estimated population-level parameters consisted of 4000 posterior samples (4 chains, 1000 samples each), returned by the No-U-Turn sampler (NUTS) \citep{hoffman_no-u-turn_2011}. A total of 16 participants were simulated conditioned on estimated population-level parameters--resulting in 4000 distinct draws, each consisting of 16 participants. The threshold values for these draws were used as ground truth for a comparative analysis in Results \ref{res-accuracy} (Fig. \ref{fig-accuracy-power}e,f).

The default hierarchical Bayesian model (HB, Supplementary Fig. \ref{sufig-meth_default}g,h) was benchmarked against three non-hierarchical models to assess the effect of partial pooling across participants. These models included a non-hierarchical Bayesian (nHB) model, implemented equivalently to the HB model using the same priors, except without pooling; maximum likelihood (ML) model implemented with uniform priors for the participant-level parameters; and the Nelder-Mead (NM) method, which utilized the SciPy  \citep{virtanen_scipy_2020} library to minimize the residual sum of squares between the data points and the recruitment curve fit (cost function). The NM method was re-initialized 100 times with different starting points to avoid local minima and the threshold point estimate was chosen based on the minimum cost function. For the HB, nHB, and ML models, point threshold estimates were calculated using the mean of threshold posterior. These point estimates were used to compute mean absolute error from ground truth thresholds. Figure \ref{fig-accuracy-power}e consisted of 48 equispaced stimulation intensities between 0--100\% MSO (in Supplementary Fig. \ref{sufig-meth_default}g,h $n(p) = 48\;\forall p$). Figure \ref{fig-accuracy-power}f consisted of exactly the first 8 participants of each draw ($N_P = 8$). Both analyses involved single repetition per intensity and were repeated for 4000 draws.

The mean absolute error was calculated as follows: let $\{a_1^d, a_2^d, \ldots, a_{16}^d\}$ be the true thresholds for the sixteen participants of the $d$-th draw, and $\{\hat{a}_1^d, \hat{a}_2^d, \ldots, \hat{a}_{n}^d\}$ be the corresponding point estimates of a model for the first $n\in\{1, 2, \ldots 16\}$ participants. Then, the error for $n$ participants of the $d$-th draw is given by $e_{n, d} = \frac1n \sum_{p=1}^{n}\lvert a_p^d - \hat{a}_p^d \rvert$. Finally, the error for $n$ participants (Fig. \ref{fig-accuracy-power}e,f) across all draws is given by $e_{n} = \frac1{4000}\sum_{d=1}^{4000}e_{n, d}$, which is the sample mean of $\{e_{n, 1}, e_{n, 2} \ldots e_{n, 4000}\}$. The error bars (Fig. \ref{fig-accuracy-power}e,f) represent the standard error of $e_{n}$ given by $\text{SE}_{e_n} = \frac{\sigma_{e_n}}{\sqrt{4000}}$, where $\sigma_{e_n} = \sqrt{\sum_{d=1}^{4000}\left(e_{n,d} - e_n\right)^2 / \left(4000 - 1\right)}$ is the sample standard deviation.

\subsubsection{Bayesian estimation for detecting shift in threshold}\label{meth-power}
\noindent In Results \ref{res-power} (Fig. \ref{fig-accuracy-power}g,h), we simulated 4000 draws of 20 participants. The parameters of both pre- and post-intervention phases, except for the post-intervention thresholds, were simulated by conditioning the model (Supplementary Fig. \ref{sufig-meth_default}g,h) on the estimated population-level parameters. The post-intervention thresholds were obtained by subtracting values from the pre-intervention thresholds, where the differences were simulated from a normal distribution $\mathcal{N}\left(\mu=-5, \sigma=2.5\right)$ for a negative shift, and $\mathcal{N}\left(0, 2.5\right)$ for no shift. This assumed that the intervention did not alter the population-level distribution of any parameter other than the threshold.

The null hypothesis assumed no shift from pre- to post-intervention, whereas the alternative hypothesis posited a negative shift. Supplementary Fig. \ref{sufig-power} specifies the within-participant comparison model (HB, Methods \ref{meth-within}) that was benchmarked against non-hierarchical models: non-hierarchical Bayesian (nHB), maximum likelihood (ML), and Nelder-Mead (NM). For the HB model, the null hypothesis was rejected if the 95\% HDI of the population-level location hyperparameter $\left(\mu_{a_\Delta}\right)$ was entirely left of zero; otherwise, the null hypothesis was not rejected. For the nHB, ML, and NM models, a one-sided Wilcoxon signed-rank test on the point estimates of pairwise threshold differences was applied. The significance level was set at 5\% and the null hypothesis was rejected if the $p$-value was below 0.05. A $t$-test wasn't applicable due the non-normality of estimated differences indicated by Shapiro-Wilk test.

The 4000 draws were shuffled and both analyses (Fig. \ref{fig-accuracy-power}g,h) were repeated for the first 2000 draws. Note that these couldn't be repeated for all draws since subtracting values from the pre-intervention thresholds would sometimes result in negative post-intervention thresholds for some participants. A draw was discarded if any participant had a negative post-intervention threshold, resulting in 96.88\% and 99.72\% valid draws when differences were simulated from $\mathcal{N}\left(-5, 2.5\right)$ and $\mathcal{N}\left(0, 2.5\right)$ respectively. The decision to repeat for the first 2000 shuffled draws was made without peeking at the results.

The true and false positive rates were calculated as follows: let $H_0$ be the null hypothesis (no shift), $H_1$ be the alternative hypothesis (negative shift). We define the indicator variable $\mathbf{1}_{n,d}$ which evaluates to 1 if the statistical test (95\% HDI testing for the HB model, and one-sided signed-rank test for the non-hierarchical models) rejects $H_0$ based on the first $n$ participants of the $d$-th draw, and 0 otherwise. Define $\pi_{n} = \frac{1}{2000}\sum_{d=1}^{2000} \mathbf{1}_{n,d}$, which is the sample mean of the set of binary values $\left\{\mathbf{1}_{n,1}, \mathbf{1}_{n,2} \ldots \mathbf{1}_{n,2000}\right\}$. When the differences come from $\mathcal{N}\left(-5, 2.5\right)$, the null hypothesis is false and the true positive rate (Fig. \ref{fig-accuracy-power}g) is given by $\pi_{n}$. When the differences come from $\mathcal{N}\left(0, 2.5\right)$, the null hypothesis holds and the false positive rate (Fig. \ref{fig-accuracy-power}h) is given by this same quantity. The error bars represent the standard error of $\pi_n$.

\subsection{Choice of recruitment curve function}\label{meth-choice}
\noindent In Results \ref{res-choice} (Fig. \ref{fig-cross_validation_functional}e--g), we used the default model structure (Methods \ref{meth-default}) for cross-validation \citep{vehtari_practical_2017}. Supplementary Fig. \ref{sufig-cv_rectified_logistic} specifies the default rectified-logistic model for rat epidural SCS, human TMS and human epidural SCS datasets. The rectified-logistic participant-level parameters were replaced with those of logistic-5, logistic-4 and rectified-linear. Supplementary Fig. \ref{sufig-cv_logistic5}--\ref{sufig-cv_rectified_linear} specify the models for each function on all datasets. For the rat epidural SCS data, 150 recruitment curves were fit simultaneously on six muscles: abductor digiti minimi (ADM), biceps, deltoid, extensor carpi radialis longus (ECR), flexor carpi radialis (FCR), and triceps--for a total of 900 curves. For the human TMS data, 27 curves  were fit simultaneously on six muscles: ADM, abductor pollicis brevis (APB), biceps, ECR, FCR, and triceps--for a total of 162 curves. For the human SCS data, 26 curves were fit simultaneously on four muscles: ADM, APB, biceps, and triceps--for a total of 104 curves. An arviz \citep{kumar_arviz_2019} implementation of cross-validation \citep{vehtari_practical_2017} was used to compute expected log-pointwise predictive density (ELPD) scores and pairwise differences from the best-ranked model.

\bigskip\noindent In Results \ref{res-robustness} (Fig. \ref{fig-cross_validation_mixture}c), the default rectified-logistic function with Gamma observation model (Methods \ref{meth-default}, Supplementary Fig. \ref{sufig-cv_rectified_logistic}, Eq. \ref{eq416}--\ref{eq418}) was compared to its mixture extension (Methods \ref{meth-mixture}, Eq. \ref{eq425}--\ref{eq428}). Supplementary Fig. \ref{sufig-cv_rectified_logistic_mixture} specifies the mixture model on all datasets.

\subsection{Common use cases}\label{meth-uses}
\subsubsection{Within-participant comparison}\label{meth-use-within}
\noindent In Results \ref{res-within} (Fig. \ref{fig-use_cases_within}), we used the mixture extension (Methods \ref{meth-mixture}) of the within-participant comparison model (Methods \ref{meth-within}) to estimate threshold differences between midline and lateral stimulation. Supplementary Fig. \ref{sufig-use_case_within} specifies the model used for this analysis. Here, $c=1$ and $c=2$ represent lateral and midline stimulation respectively. Figure \ref{fig-use_cases_within}b displays the 95\% HDI of the location hyperparameter $\left(\mu_{a_\Delta}\right)$ for each muscle. All muscles of the arm and hand (biceps, triceps, APB, and ADM) that were consistently recorded from 13 participants were analyzed.

\subsubsection{Between-groups comparison}\label{meth-use-group}
\noindent In Results \ref{res-between} (Fig. \ref{fig-use_cases_between}), we used the mixture extension (Methods \ref{meth-mixture}) of the between-groups comparison model (Methods \ref{meth-between}) to estimate threshold differences between groups of uninjured participants and participants living with spinal cord injury. Supplementary Fig. \ref{sufig-use_case_between} specifies the model used for this analysis. Figure \ref{fig-use_cases_between}c displays the 95\% HDI of posterior difference $\left({\mu_a}^{g_1,m} - {\mu_a}^{g_2,m}\right)$ for each muscle, where $g_1$ and $g_2$ represent groups of SCI and uninjured participants respectively. We analyzed data from 13 SCI and 14 uninjured participants for all muscles of the hand and wrist (APB, ADM, ECR, and FCR).

\subsubsection{Optimizing experimental design}\label{meth-use-optimization}
Additionally, in Results \ref{res-reps} (Supplementary Fig. \ref{sufig-repetitions}) the effect of single versus multiple repetitions per pulse on threshold estimation accuracy was assessed. For the first 8 participants of the 4000 draws (Methods \ref{meth-accuracy}), 8 observations per stimulation intensity were simulated, at a total of 64 equispaced intensities between 0--100\% MSO. With repetition counts $r \in \{1, 4, 8\}$ and total intensities (including repetitions) $T\in\{32, 40, 48, 56, 64\}$ ($n(p) = T\;\forall p$), the number of unique intensities tested was given by $T / r$, which were subsampled from the initial 64 equispaced intensities. Supplementary Fig. \ref{sufig-meth_default}g,h specifies the model that was used to estimate the thresholds. This result was repeated for 4000 draws. The errors were calculated similar to Methods \ref{meth-accuracy}.

\subsection{Statistics \& reproducibility}\label{meth-reproducibility}
All models were fit with NumPyro's \citep{phan2019composable,bingham2019pyro} implementation of No-U-Turn (NUTS) \citep{hoffman_no-u-turn_2011} sampler. The code to reproduce the presented analyses is available on GitHub (see Code availability \ref{code-availability}).

\subsection{Human TMS and SCS data}\label{meth-human_data}
All procedures were reviewed and approved by the Institutional Review Board (IRB) of James J. Peters Veterans Affairs Medical Center (JJP VAMC); Weill Cornell Medicine (WCM-IRB, 1806019336); and Columbia University Irving Medical Centre (IRB 2, protocol AAAT6563). The study was pre-registered at clinicaltrials.gov (NCT05163639). Written informed consent was obtained prior to study enrollment, and all experimental procedures were conducted in compliance with institutional and governmental regulations guiding ethical principles for participation of human volunteers. The goal of the study consisted of assessing the synergistic \citep{mcintosh_timing-dependent_2024} and plasticity inducing effects of combining brain (TMS or transcranial electrical stimulation) and spinal cord (transcutaneous or epidural) stimulation in uninjured and SCI participants. TMS and epidural SCS recruitment curve data was extracted from one session per participant in the presented analysis.

\subsubsection{Human TMS data}\label{meth-human_tms}
Individuals between the ages of 18 and 80 without neurological injury (uninjured volunteers) and individuals with chronic ($>$ 1 year) cervical SCI, were eligible for recruitment. SCI participants required partially retained motor hand function, scoring 1-4 (out of 5) on manual muscle testing, with detectable TMS-evoked MEPs (greater than 50 $\mu$V) of the left or right target muscle. Recruitment curve data from 13 individuals living with SCI and 14 individuals with no neurological deficits available at the time of this study were used for analysis. SCI motor level and impairment severity were determined by clinical examination according to the International Standards for the Neurological Classification of SCI (ISNCSCI). Surface electromyography (EMG) preamplifiers were placed bilaterally over the APB, FDI, ADM, FCR, ECR, biceps brachii short head (which we refer to as biceps), triceps brachii long head (which we refer to as triceps), and tibialis anterior (TA) muscle in a belly-tendon montage, as described previously \citep{wu_posteroanterior_2020}. EMG signals were bandpass filtered between 15 and 2000 Hz, and sampled at 5000 Hz via an MA400 EMG system (Motion Lab Systems Inc., Louisiana, USA).

TMS was delivered with a MagPro X100 system (MagVenture Inc., Georgia, USA) with 80mm winged coil (D-B80; MagVenture Inc.) placed over the hand motor cortex (M1) hotspot for optimal response in the target muscle.  Electromagnetic stimulation was delivered as a single biphasic sinusoidal (anodic-first; 0.5 ms/phase) pulse, triggered by an analog-to-digital acquisition system (NI USB-6363).

Recruitment curves were assembled via delivery of TMS pulses of varying intensities in pseudorandom order ranging from subthreshold to 200\% or more of threshold using customized LabVIEW algorithms. In each participant, a total of 61.4 $\pm$ 14.0 stimulation pulses between 26.0 $\pm$ 13.3\% to 81.1 $\pm$ 16.1\% MSO. In 16 participants (SCI $n = 7$) the total number of pulses were divided into 7-8 repetitions per stimulation intensity. In the remaining 11 participants (SCI $n = 6$), the stimulation protocol was changed so that each stimulation trial had a unique intensity based on the preliminary development of our hbMEP approach (see Supplementary Fig. \ref{sufig-repetitions}). MEPs in triceps, biceps, ECR, FCR, APB and ADM contralateral to the site of stimulation were quantified as peak-to-peak in an 83.5 ms window starting at 6.5 ms after the start of the first stimulation pulse. Due to temporal jitter in the recording system for the triceps muscle, the starting point of the window was increased to 10.6 $\pm$ 2.1 ms in order to avoid stimulation artifacts.

\subsubsection{Human epidural SCS data}\label{meth-human_scs}
Detailed protocols can be found in McIntosh et al., 2023 \citep{mcintosh_intraoperative_2023}, relevant sections are reproduced here. Participants were adult patients with cervical spondylotic myelopathy and/or multilevel foraminal stenosis requiring surgical intervention. Patients were enrolled from the clinical practices of the spine surgeons participating in the study.

Epidural electrodes were used for stimulation during clinically indicated surgeries, with EMG recordings taken from muscles selected as per standard of care. Recordings were made at a sampling rate between 6 kHz and 10.4 kHz, band-pass filtered between 10 Hz and 2,000 Hz. A three-pulse train was used for epidural spinal cord stimulation to reduce the necessary intensity to evoke an MEP.

In 13 participants, stimulation was applied at the most caudal exposed segment at midline and lateral locations to compare recruitment curves. A handheld double-ball tip epidural electrode was positioned at midline, in line with the dorsal root entry zone. Stimulation intensity was incremented from 0 up to 8 mA to assess the activation threshold and estimate the subsequent recruitment curve (minimum 5 MEPs per stimulation intensity). The experiments proceeded by repeating the stimulation intensity ramp and fixed-intensity stimulation at the equivalent lateral site.
MEPs were quantified in biceps, triceps, APB and ADM ipsilateral to the side of stimulation with the rectified AUC calculated in a window between 6.5 ms and 75 ms after the start of the first stimulation pulse.

\subsection{Rat epidural SCS data}\label{meth-rat_data}
Eight Sprague Dawley rats were used in this study for a terminal physiology experiment. All procedures were conducted in compliance with the guidelines of the Institutional Animal Care and Use Committee at Columbia University in New York, NY, and followed aseptic techniques. Detailed methodology of the protocol used can be found in Mishra et al., 2017 \citep{mishra_paired_2017} and Pal et al., 2022 \citep{pal_spinal_2022}.

EMG activity was recorded from 8 different muscles: left ECR, FCR, biceps, triceps, ADM, deltoid, biceps femoris, and right biceps. Flexible, braided stainless steel wires were employed for EMG recording. A hole was drilled into the skull between the eyes, and the ground screw electrode was inserted into the hole. EMG and ground electrodes were soldered to the connector and covered with epoxy to ensure insulation. Subsequently, the connector was attached to the recording system.

After placing the EMG electrodes, the animal's head was fixed and the T1 spinous process was clamped to stabilize the spine. The C4 spinal cord was exposed by laminectomy. Custom designed electrode arrays \citep{garcia-sandoval_chronic_2018} were placed in the dorsal epidural space in the midline of the spinal cord over the cervical enlargement (C5-C8). The arrays consisted of 12 electrodes arranged in a 4 by 3 configuration, with the outer columns aligned to each of the C5-C8 dorsal root entry zones and the central columns aligned to the spinal cord midline. The muscles over the spinal cord were brought back together to prevent temperature loss and reduce dryness around the spinal cord opening. Omnetics connectors (Omnetics Connector Corp.; Minneapolis, USA) for the spinal array and EMG wires were mounted on the skull for stimulation and recording.

Connectors for the spinal cord array and EMG wires were attached to a headstage ZIF-clip (Tucker-Davis Technologies; Florida, USA) via omnetics connectors. Raw signals were sampled at 10 kHz. ZIF-connectors were used to interface the implanted electrodes with a PZ5 amplifier (Tucker-Davis Technologies) in turn connected to a real-time RZ2 signal processing system (Tucker-Davis Technologies).

A 16-channel IZ2H constant current stimulator (Tucker-Davis Technologies), controlled via custom Matlab (R2022a) scripts, delivered biphasic single-pulse stimulation of 200 $\mu$s every 2 seconds, with intensities linearly increased from 0 to an average of 325 $\pm$ 88.6 $\mu$A across 51 steps. Stimulation patterns were randomly applied across 21 spatial combinations on the left side and midline of the array, excluding high impedance electrodes from testing, resulting in an average of 18.8 $\pm$ 3.2 combinations tested per rat. EMG signals were high-pass filtered using a 20 Hz cutoff 10th order IIR filter. MEPs were quantified in biceps, triceps, ECR, FCR, APB and ADM ipsilateral to the side of stimulation by calculating the AUC within a 1.5 to 10 ms window post-stimulation onset.

%% file: sections/additional.tex
\section{Additional information}
\subsection{Competing interests}
Jason B. Carmel is a Founder and stock holder in BackStop Neural and a scientific advisor and stockholder in SharperSense. He has received honoraria from Pacira, Motric Bio, and Restorative Therapeutics. Noam Y. Harel is a consultant for RubiconMD. Michael S. Virk has been a consultant and has received honorarium from Depuy Synthes and BrainLab Inc; he is on the Medical Advisory Board and owns stock with OnPoint Surgical. The other authors have nothing to disclose.

\subsection{Funding}
This research was supported by the National Institutes of Health (R01NS124224, R01NS115470); and the Travis Roy Foundation Boston, MA (Investigator Initiated).

\subsection{Acknowledgements}
We thank A. Gelman and S. Fara for helpful feedback and comments on the manuscript.

\bigskip\noindent Human TMS data: We thank J. A. Goldsmith, Y. Wu, K. S. Holbrook, M. Liu, T. Magno, G. Famodimu, G. A. Mendez, J. M. Robbins, L. E. Kinne, and A. Villaroel-Sanchez at the James J. Peters VAMC for helping obtain human TMS data.

\bigskip\noindent Human epidural SCS data: We thank neurologists A. Mendiratta (Weill Cornell Medicine), S. C. Karceski, and M. Bell (The Och Spine Hospital At New York Presbyterian Hospital) and intraoperative monitoring technologists E. Thuet (The Och Spine Hospital At New York Presbyterian Hospital), E. Shelkov (Weill Cornell Medicine) and their teams including O. Modik and N. Patel (Weill Cornell Medicine). We also thank M. Vulapalli, C. Mykolajtchuk, and M. Michael (Weill Cornell Medicine) for help in administrative matters.

\subsection{Author contributions}
Conceptualization: JRM; Data Curation: LMM, ASA, NYH, JRM; Formal Analysis: VT, JRM; Funding Acquisition: MSV, NYH, JBC, JRM; Investigation: VT, LMM, ASA, CM, MSV, NYH, JBC, JRM; Methodology: VT, JRM; Project Administration: MSV, NYH, JBC, JRM; Resources: CM, MSV, NYH, JBC; Software: VT, JRM; Supervision: JBC, JRM; Validation: VT, JBC, JRM; Visualization: VT, JBC, JRM; Writing - Original Draft: VT, JRM; Writing - Review \& Editing: VT, LMM, ASA, CM, MSV, NYH, JBC, JRM.

\subsection{Data availability}
The datasets used in this study are available from the corresponding authors upon reasonable request.

\subsection{Code availability}\label{code-availability}
The hbMEP Python package was used to perform all analyses in the paper. The code, together with tutorials, is available at https://github.com/hbmep/hbmep. Code to reproduce the presented analyses is available at https://github.com/hbmep/hbmep-paper.

%% file: sections/supplemental.tex
\setcounter{figure}{0}
\renewcommand{\thefigure}{S\arabic{figure}}
\renewcommand{\theHfigure}{S\arabic{figure}}
\renewcommand{\figurename}{Supplementary Fig.}

\section*{Supplementary information}
\begin{figure}[!htbp]
    \centering
    \includegraphics[width=.6\textwidth]{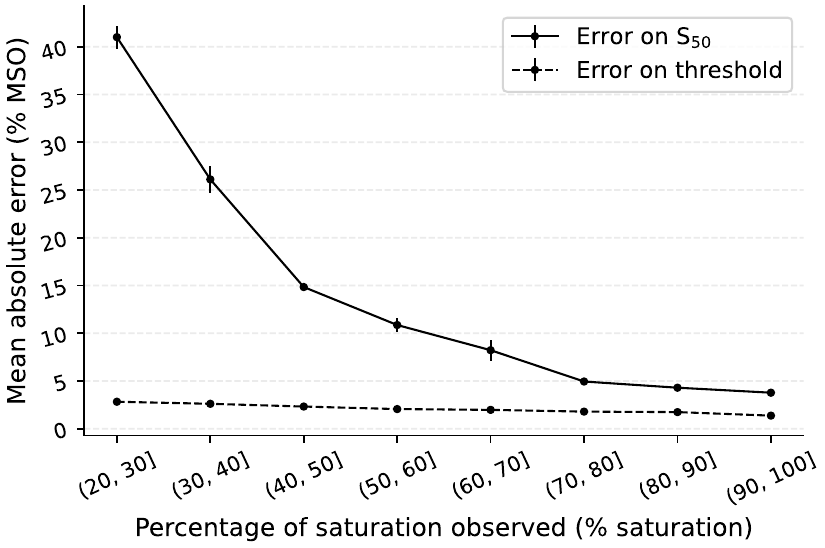}
    \caption{
        \textbf{Accurate threshold estimation is independent of saturation observed and requires less experimental burden on participants, as opposed to S$_{\mathbf{50}}$ estimation.
        } Abscissa bin $(a, b]$ represents percentage of saturation observed that is $> a$ and $\leq b$.
    }\label{sufig-saturation}
\end{figure}

\begin{figure}[!htbp]
    \centering
    \includegraphics[width=.6\textwidth]{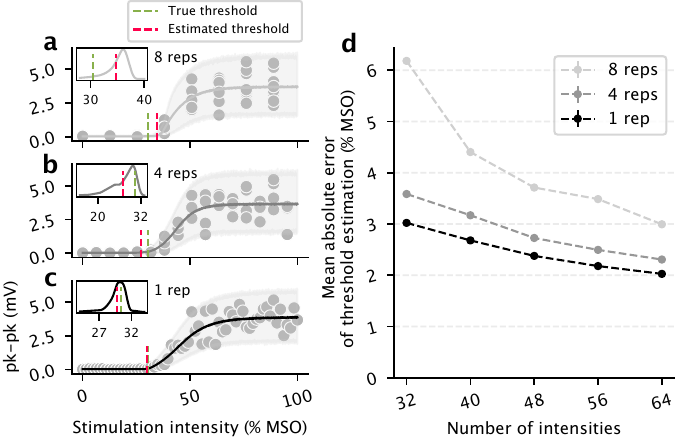}
    \caption{
        \textbf{Efficient threshold estimation is performed by sampling data evenly without repetition.} Example fits on a simulated participant with total of 64 equispaced intensities and \textbf{(a)} 8 repetitions, \textbf{(b)} 4 repetitions, \textbf{(c)} 1 repetition per intensity. \textbf{(d)} Single repetition sampling produces the lowest error regardless of the number of intensities.
    }\label{sufig-repetitions}
\end{figure}

\begin{figure}[t]
    \centering
    \includegraphics[width=.9\textwidth]{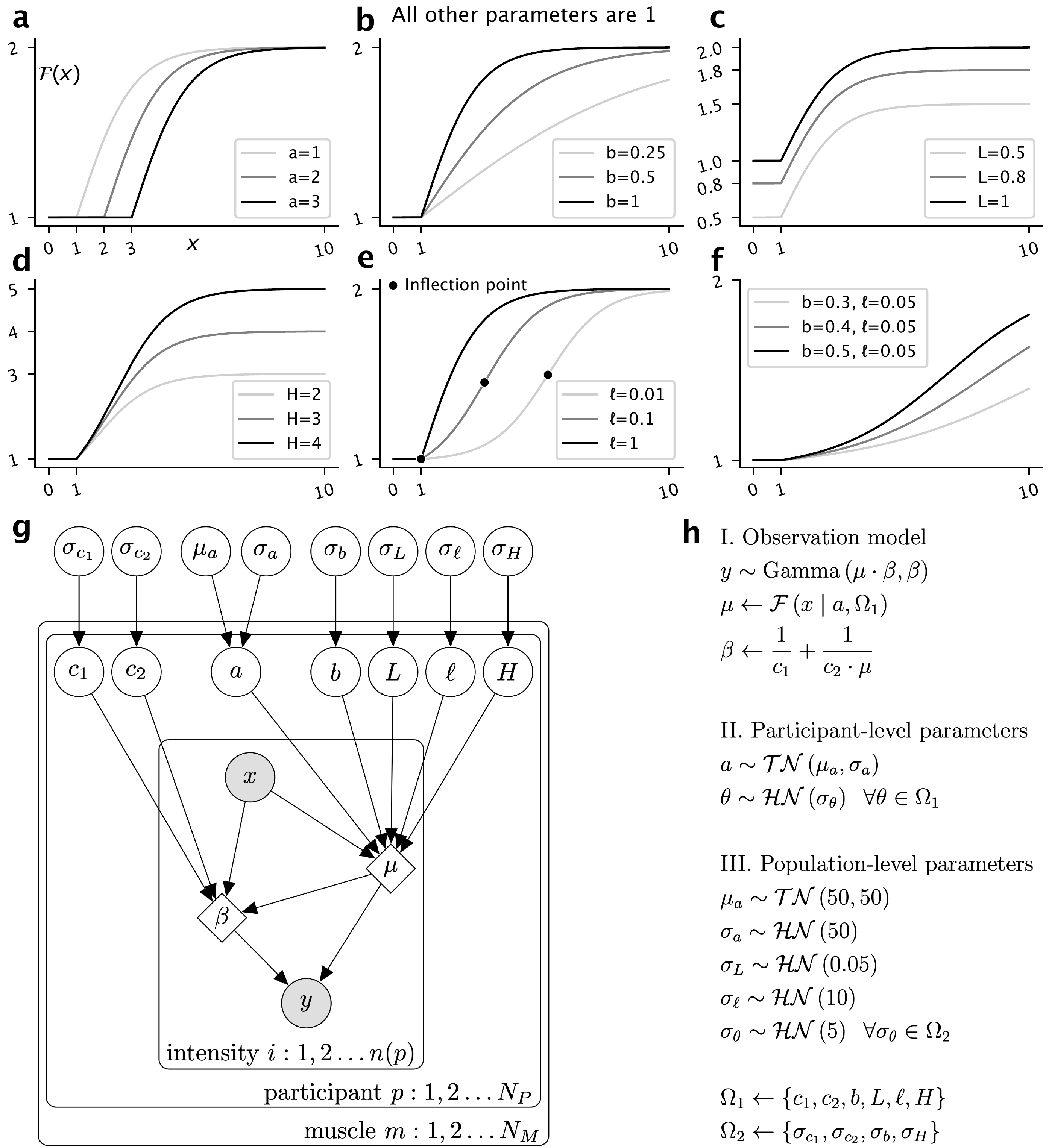}
    \caption{
        \textbf{(a--f) Effect of varying parameters of the rectified-logistic function.} \textbf{(a)} $a$ shifts the threshold. \textbf{(b)} $b$ changes the growth rate. \textbf{(c)} $L$ changes the offset MEP size. \textbf{(d)} $H$ controls the distance between offset $L$ and saturation $\left(L + H\right)$. \textbf{(e)} $\ell$ affects the location of inflection point or point of maximum gradient, whether near offset $L$ or saturation $\left(L + H\right)$. \textbf{(f)} Varying $b, \ell$ simultaneously. \textbf{(g--h) Hierarchical Bayesian model structure.} \textbf{(g)} Graphical model. The model yields parameter estimates for each participant across multiple muscles simultaneously. Circular nodes represent random variables. Filled circular nodes represent observed data. Diamonds represent deterministic variables. Arrows represent that the child node is informed by the distribution of its parent node. Plates denote re-instantiation of nodes. \textbf{(h)} Bayesian model specification with participant- and population-level parameters and weakly informative hyperpriors for TMS data. $\mathcal{F}$ is the rectified-logistic function. $\text{Gamma}$ is the gamma distribution in shape-rate parametrization. $\mathcal{TN}$ is the truncated normal distribution in location-scale parametrization with left truncation at zero. $\mathcal{HN}$ is the half-normal distribution in scale parametrization.
    }\label{sufig-meth_default}
\end{figure}

\begin{figure}[!htbp]
    \centering
    \includegraphics[width=.8\textwidth]{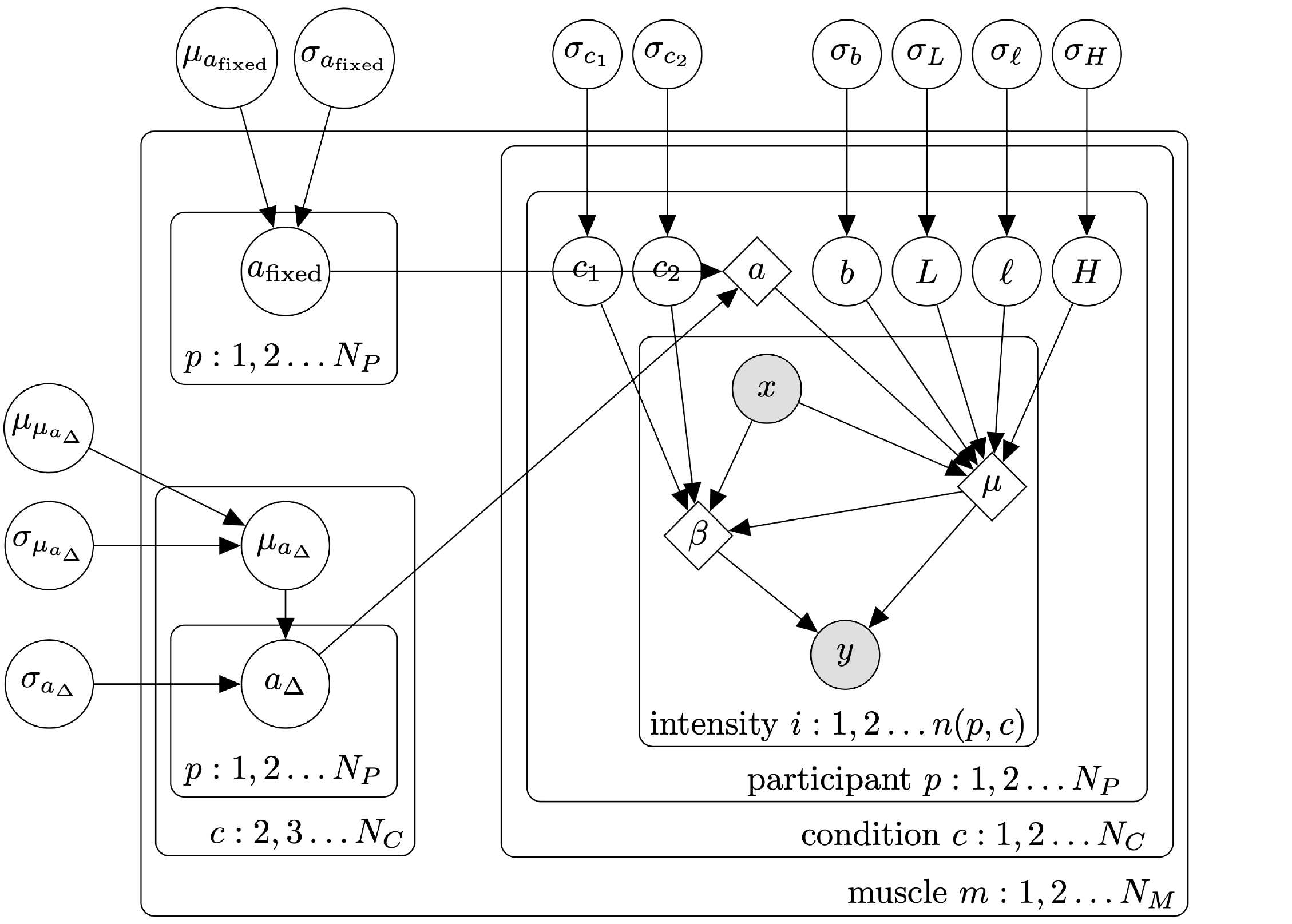}
    \caption{
        \textbf{Within-participant comparison model.} This models the within-participant differences in the threshold $(a)$ parameter, and $\mu_{a_\Delta}$ summarizes these differences across all participants.
    }\label{sufig-meth_within}
\end{figure}

\begin{figure}[!htbp]
    \centering
    \includegraphics[width=.55\textwidth]{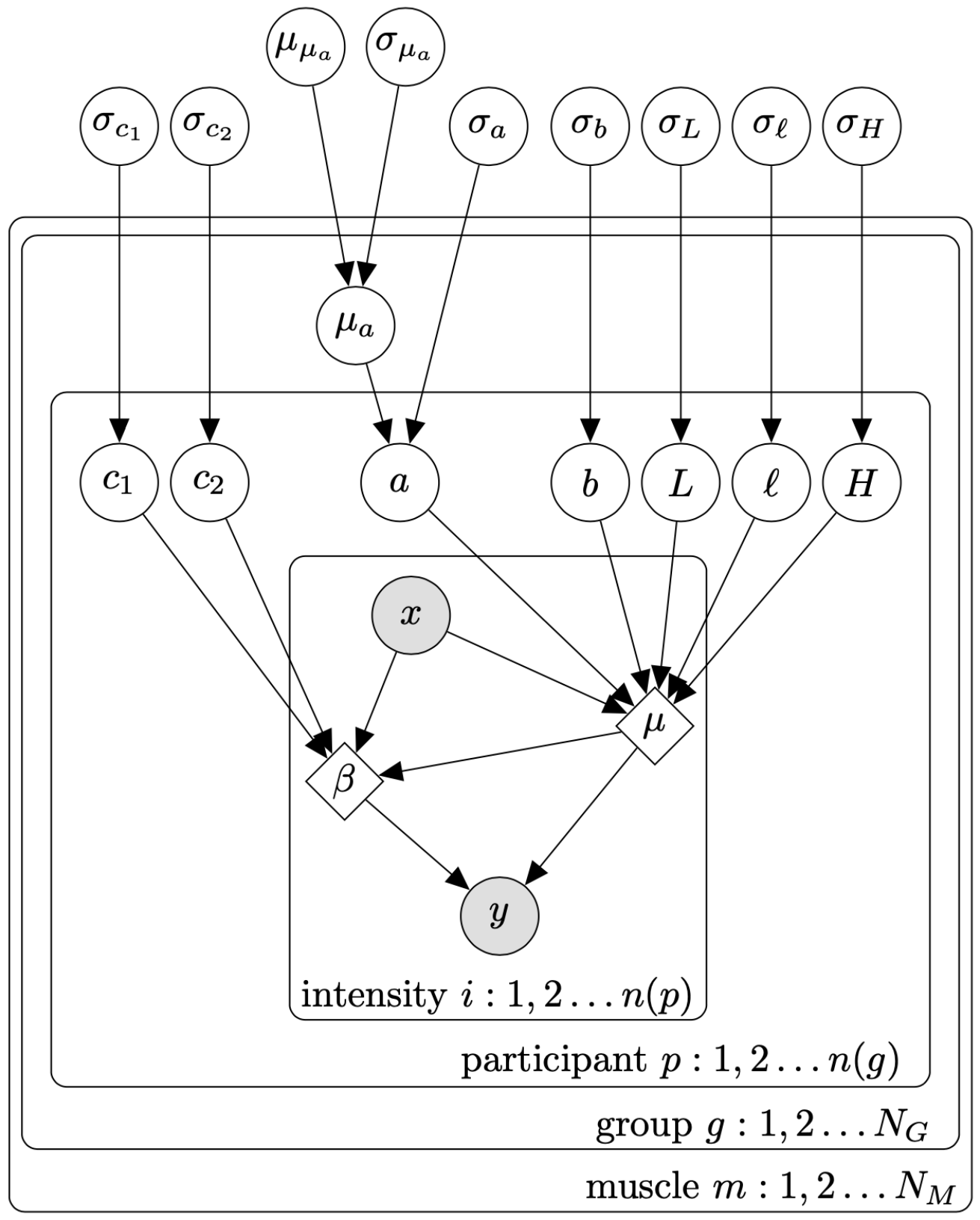}
    \caption{
        \textbf{Between-groups comparison model.} This models the distribution of threshold $(a)$ parameter for each group of participants, and the posterior difference ${\mu_{a}}^{g_{g_1},m} - {\mu_{a}}^{g_{g_2},m}$ compares the parameter between groups $g_1$ and $g_2$ at muscle $m$. The number of intensities $i$ is independent of group $g$ because any given participant $p$ belongs to exactly one group.
    }\label{sufig-meth_between}
\end{figure}

\begin{figure}[!htbp]
    \centering
    \includegraphics[width=.65\textwidth]{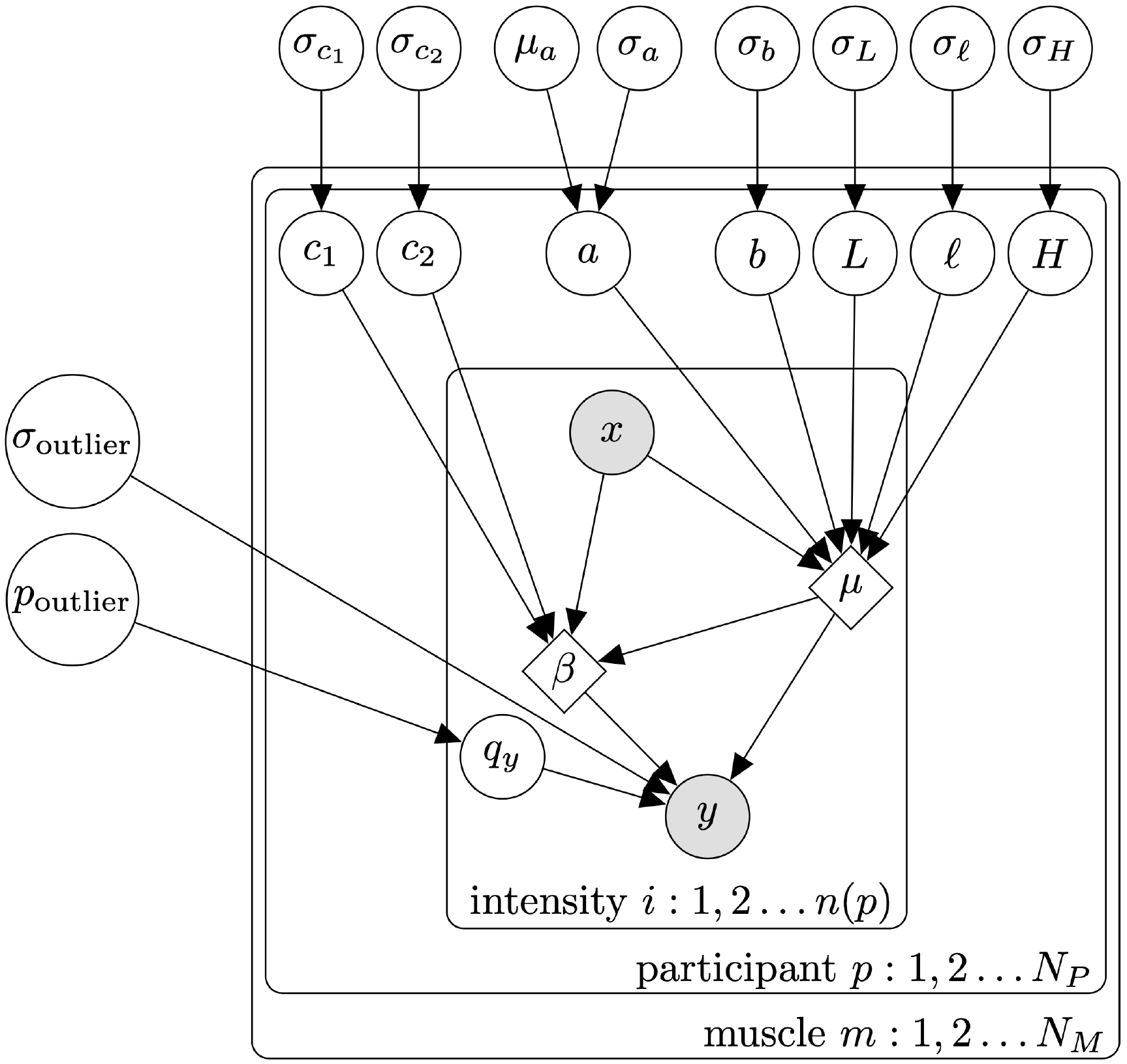}
    \caption{
        \textbf{Mixture extension of the default hierarchical Bayesian model.}
    }\label{sufig-meth_mixture}
\end{figure}

\begin{figure}[!htbp]
    \centering
    \includegraphics[width=\textwidth]{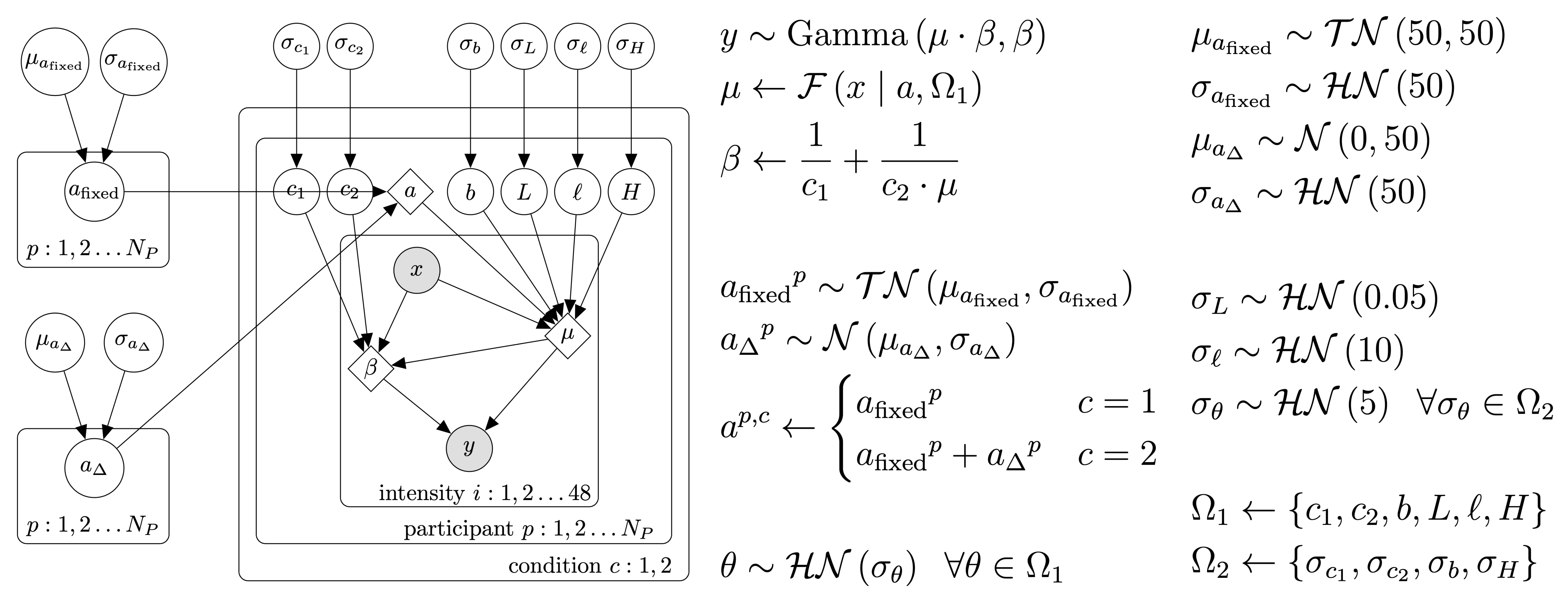}
    \caption{
        \textbf{Within-participant comparison model for detecting shift in threshold between pre- and post-intervention.} Here, $c=1$ and $c=2$ represent pre- and post-intervention conditions, respectively. A priori the model assumes no shift, as indicated by a flat prior on $\mu_{a_\Delta}$ that is symmetric about zero.
    }\label{sufig-power}
\end{figure}

\begin{figure}[!htbp]
    \centering
    \includegraphics[width=\textwidth]{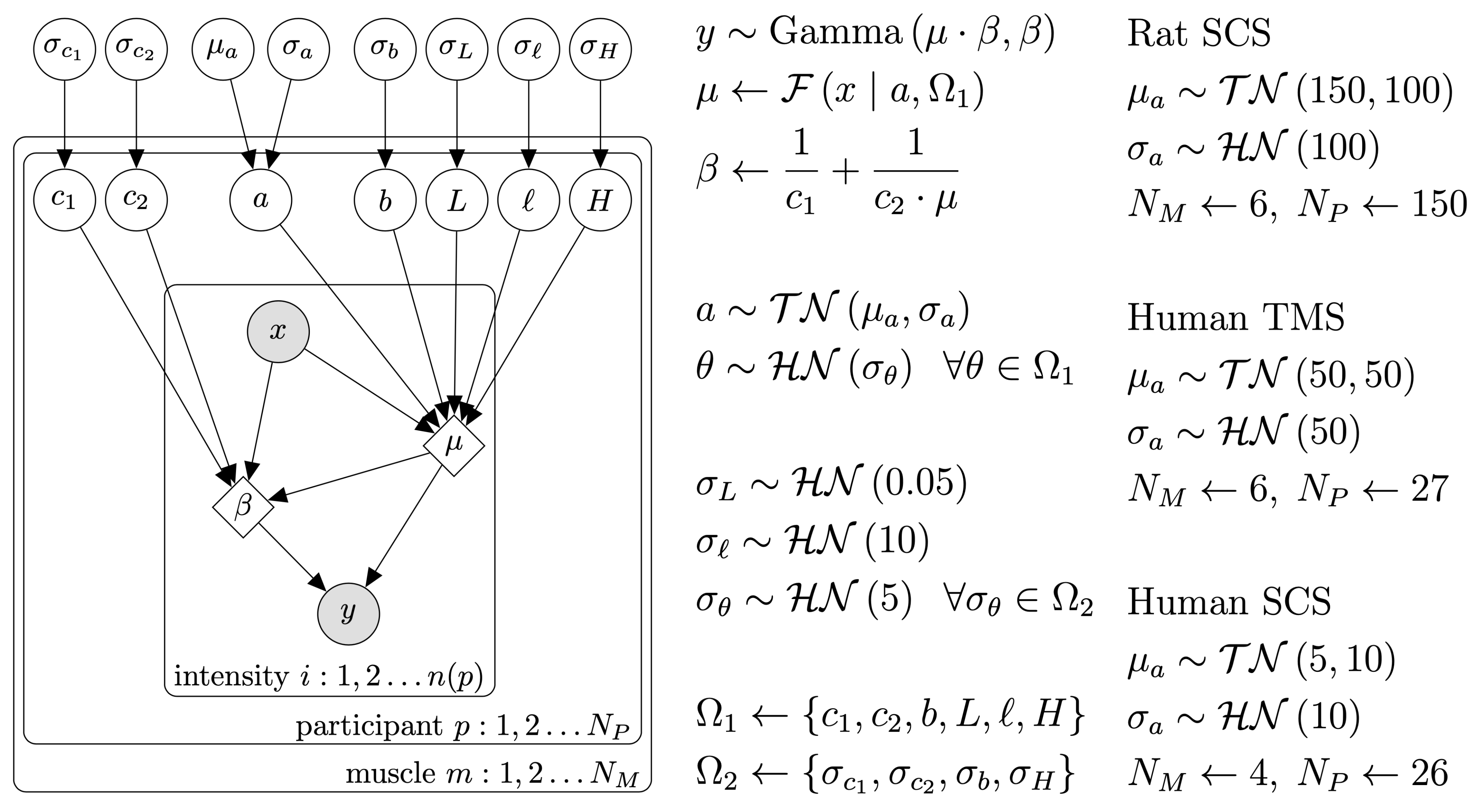}
    \caption{
        \textbf{Default rectified-logistic model for cross-validation on TMS and SCS data.}
    }\label{sufig-cv_rectified_logistic}
\end{figure}

\begin{figure}[!htbp]
    \centering
    \includegraphics[width=\textwidth]{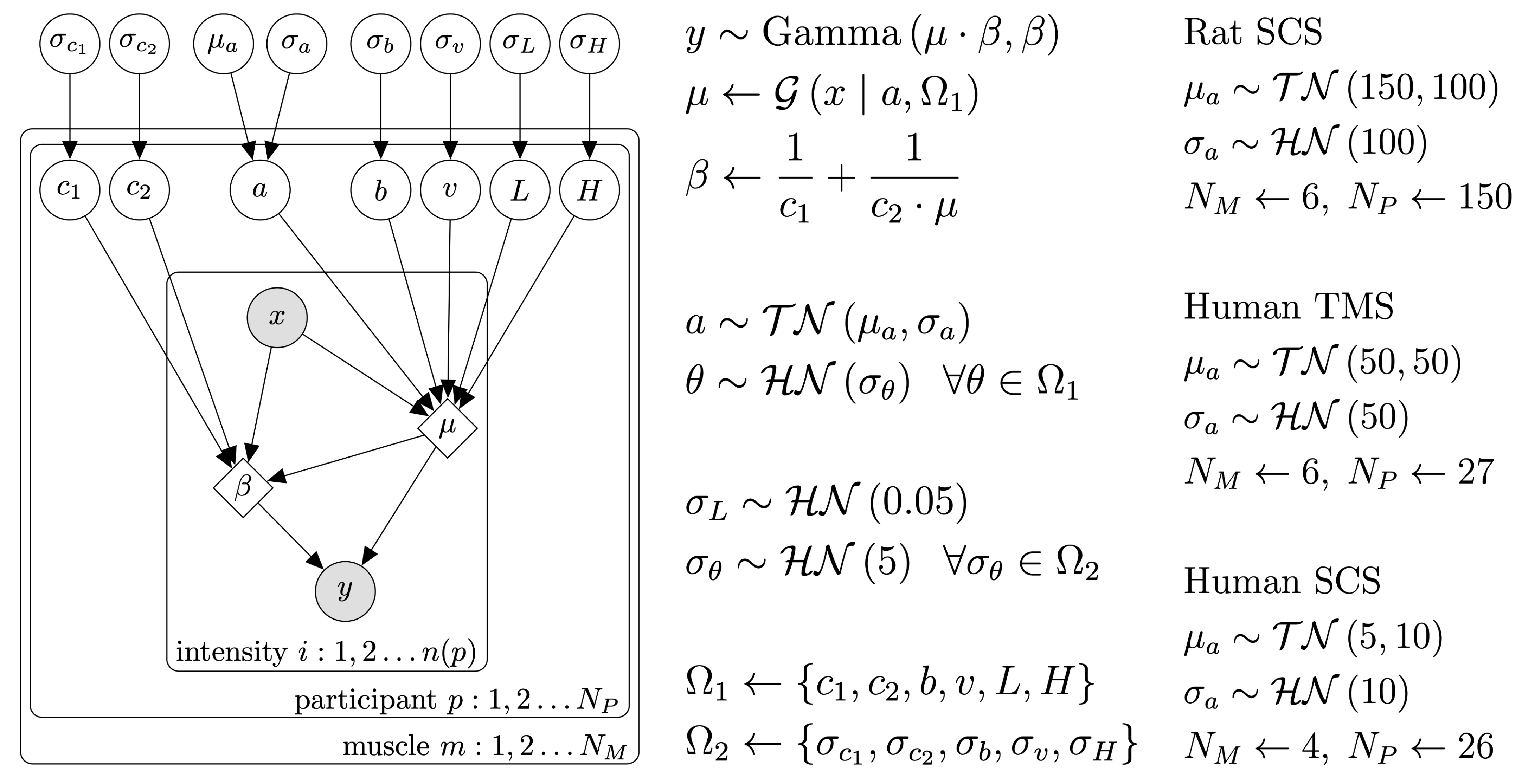}
    \caption{
        \textbf{Default logistic-5 model for cross-validation on TMS and SCS data.} $\mathcal{G}$ is the logistic-5 function.
    }\label{sufig-cv_logistic5}
\end{figure}

\begin{figure}[!htbp]
    \centering
    \includegraphics[width=\textwidth]{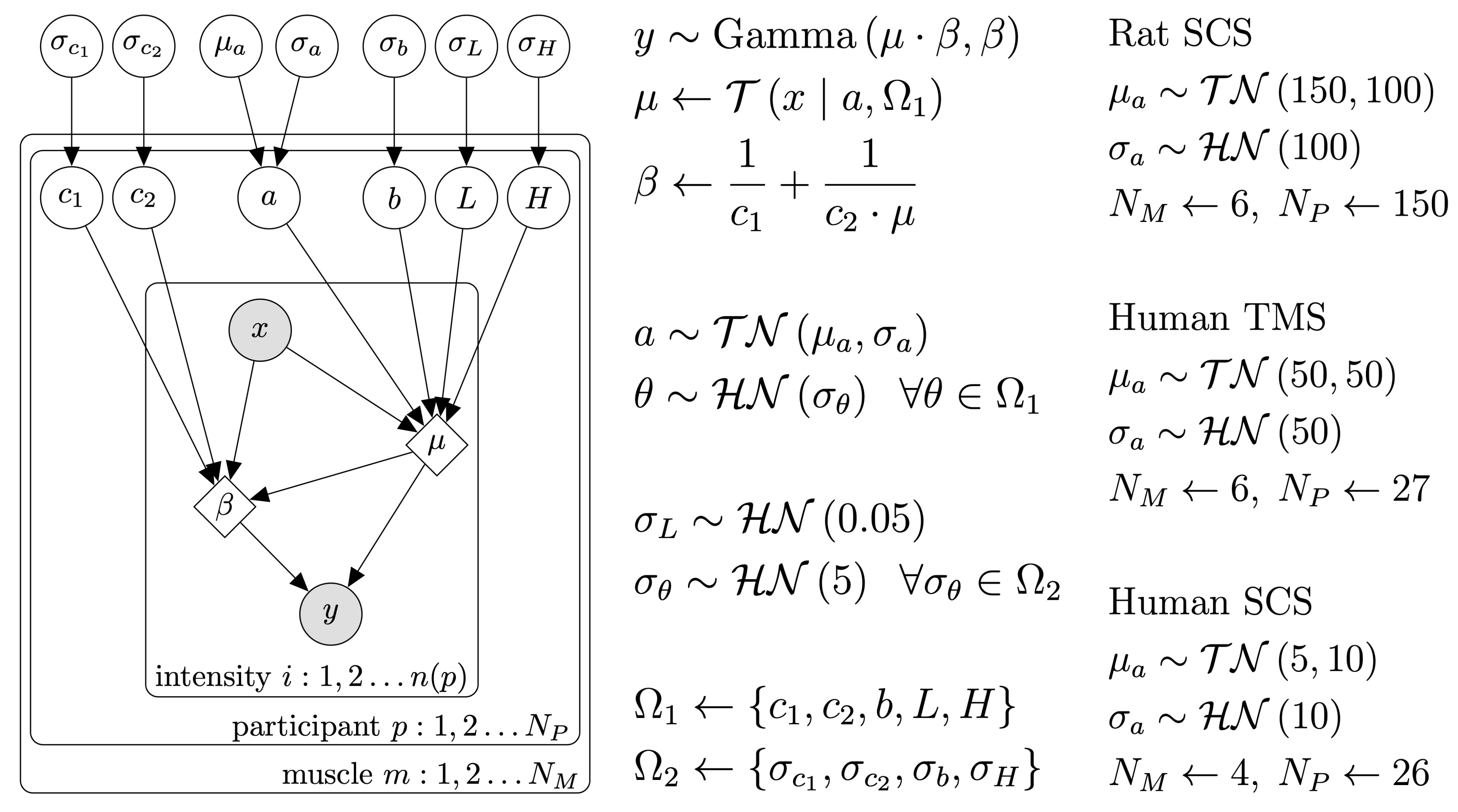}
    \caption{
        \textbf{Default logistic-4 model for cross-validation on TMS and SCS data.} $\mathcal{T}$ is the logistic-4 function.
    }\label{sufig-cv_logistic4}
\end{figure}

\begin{figure}[!htbp]
    \centering
    \includegraphics[width=\textwidth]{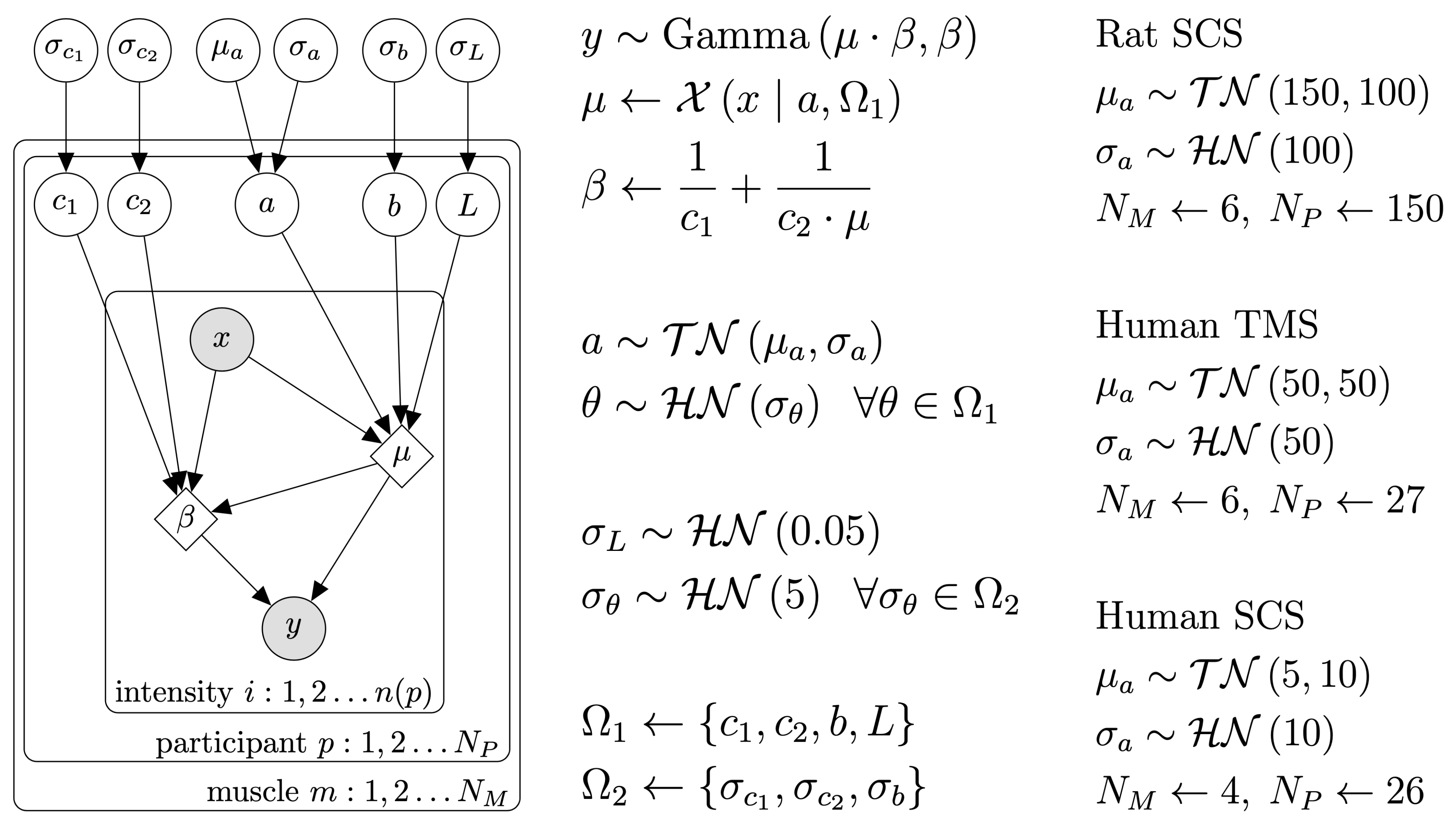}
    \caption{
        \textbf{Default rectified-linear model for cross-validation on TMS and SCS data.} $\mathcal{X}$ is the rectified-linear function.
    }\label{sufig-cv_rectified_linear}
\end{figure}

\begin{figure}[!htbp]
    \centering
    \includegraphics[width=\textwidth]{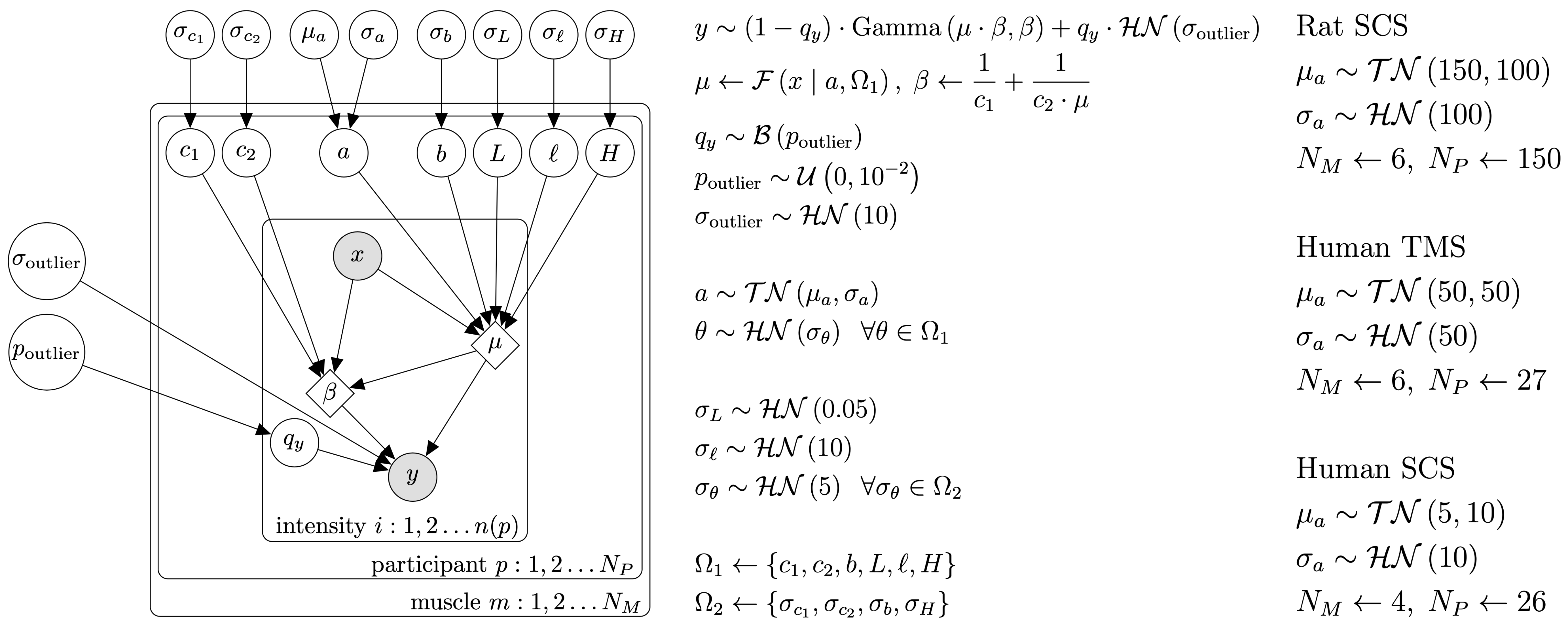}
    \caption{
        \textbf{Mixture extension of the default rectified-logistic model for cross-validation on TMS and SCS data.} $\mathcal{B}$ is the Bernoulli distribution in probability of success (success is 1, fail is 0) parametrization. $\mathcal{U}$ is the continuous uniform distribution in minimum-maximum parametrization.
    }\label{sufig-cv_rectified_logistic_mixture}
\end{figure}

\begin{figure}[!htbp]
    \centering
    \includegraphics[width=\textwidth]{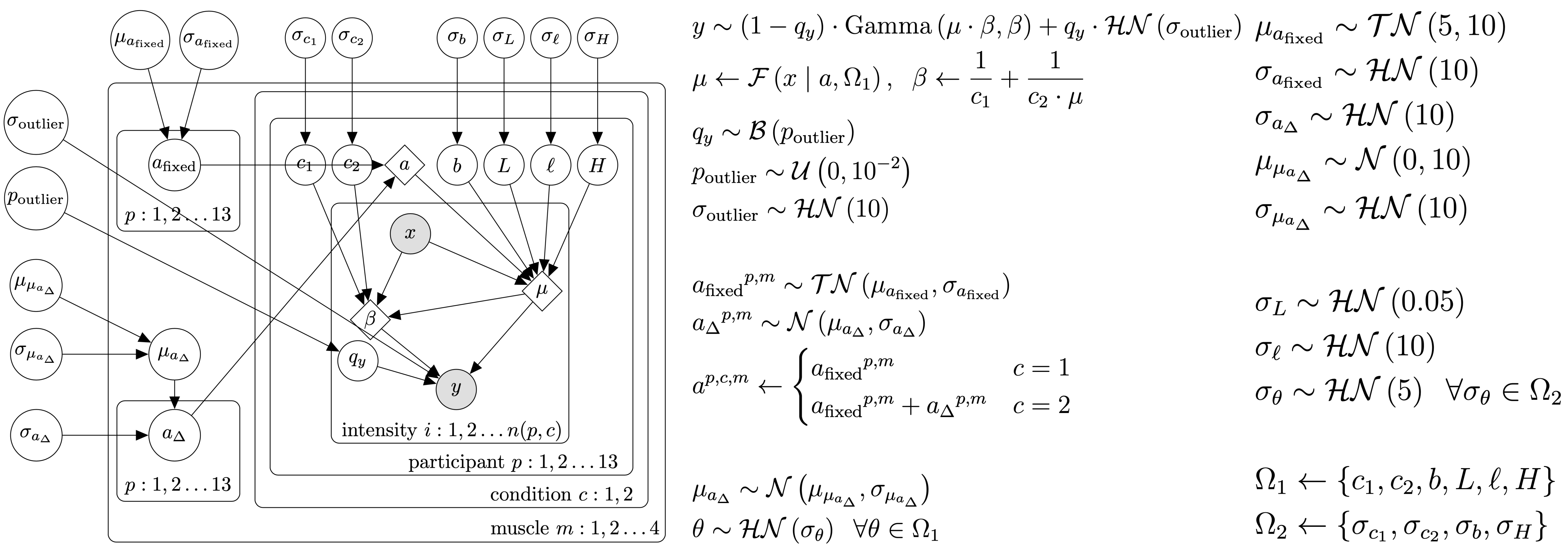}
    \caption{
        \textbf{Within-participant comparison model for comparing midline and lateral stimulation threshold.} Here, $c=1$ and $c=2$ represent lateral and midline stimulation, respectively. A priori the model assumes no shift, as indicated by a flat prior on $\mu_{\mu_{a_\Delta}}$ that is symmetric about zero.
    }\label{sufig-use_case_within}
\end{figure}

\begin{figure}[!htbp]
    \centering
    \includegraphics[width=\textwidth]{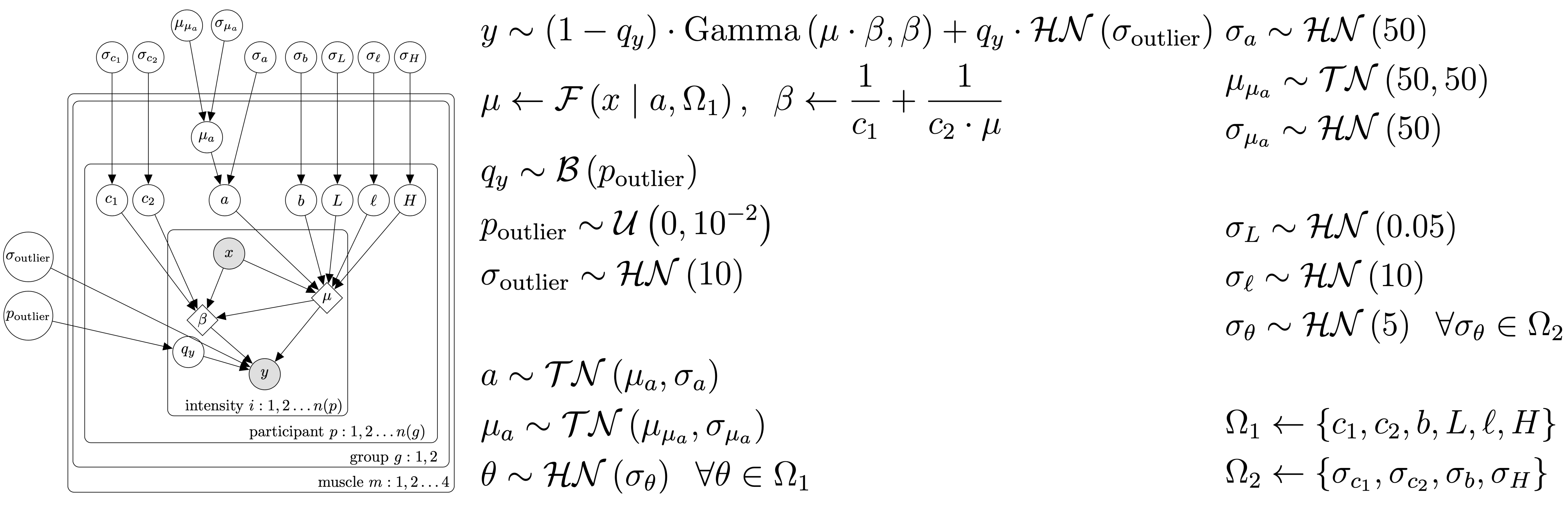}
    \caption{
        \textbf{Between-groups comparison model for comparing TMS thresholds between groups of SCI and uninjured participants.} Here, $g=1$ and $g=2$ represent SCI and uninjured groups, respectively.
    }\label{sufig-use_case_between}
\end{figure}